\documentclass[dvips]{acta}
\usepackage{supertabular,lscape,epsfig}
\usepackage{amssymb}
\usepackage{amsmath}

\SetPages{345}{362}

\SetVol{67}{2017}

\begin{document}

\begin{Titlepage}
\Title{Cool Spot and Flare Activities of a RS CVn Binary KIC 7885570}
\Author{M.~Kunt$^1$~and~H.~A.~Dal$^{1}$}
{$^1$Department of Astronomy and Space Sciences, University of Ege, Bornova,\\
35100 ~\.{I}zmir, Turkey e-mail: ali.dal@ege.edu.tr}

\Received{Received May 10, 2016}

\end{Titlepage}

\Abstract{We present here the results on the physical nature a RS CVn binary KIC 7885570 and its chromospheric activity based on the Kepler Mission data. Assuming the primary component temperature, 6530 K, the temperature of the secondary component was found to be 5732$\pm$4 K. The mass ratio of the components ($q$) was found to be 0.43$\pm$0.01, while the inclination ($i$) of the system 80$^\circ$.56$\pm$0$^\circ$.01. Additionally, the data were separated into 35 subsets to model the sinusoidal variation due to the rotational modulation, using the SPOTMODEL program, as the light curve analysis indicated the chromospherically active secondary component. It was found that there are generally two spotted areas, whose radii, longitudes and latitudes are rapidly changing, located around the latitudes of $+50^\circ$ and $+90^\circ$ on the active component. Moreover, 113 flares were detected and their parameters were computed from the available data. The One Phase Exponential Association function model was derived from the parameters of these flares. Using the regression calculations, the Plateau value was found to be 1.9815$\pm$0.1177, while the half-life value was computed as 3977.2 s. In addition, the flare frequency ($N_{1}$) - the flare number per hour, was estimated to be 0.00362 $h^{-1}$, while flare frequency ($N_{2}$) - the flare-equivalent duration emitted per hour, was computed as 0.00001. Finally, the times of eclipses were computed for 278 minima of the light curves, whose analysis indicated that the chromosphere activity nature of the system causes some effects on these minima times. Comparing the chromospheric activity patterns with the analogues of the secondary component, it is seen that the magnetic activity level is remarkably low. However, it is still at the expected level according to the B-V color index of 0.643 mag for the secondary component.}

{techniques: photometric -- methods: data analysis -- methods: statistical -- binaries: eclipsing -- stars: flare -- stars: individual (KIC 7885570)}

\section{Introduction}

About sixty five percent of stars in our Galaxy are red dwarfs. Some of them exhibit stellar spot activity or flare activity (as in the case of UV Cet stars), or both. Seventy-five percent of red dwarfs show flare activity (Rodon\'{o} 1986). It is well established that UV Cet objects are young stars on the main-sequence (Poveda et al. 1996), belonging to open clusters or associations. Strong flare activity is present in the majority of young cluster members (Mirzoyan 1990, Pigatto 1990) but the fraction of active stars decreases with the cluster age. The phenomenon can be understood as a result of decreasing of star rotation velocity with time - Skumanich's law (Skumanich 1972, Pettersen 1991, Stauffer 1991, Marcy and Chen 1992). The flare activity is related to the strength of magnetic field, which increase with rotation velocity for red dwarfs. Because the flare activity is the main source of mass loss in cool main sequence stars, the mass loss rate can be very high for stars on the Zero Age Main Sequence. In the case of UV Cet type stars, it is about $10^{-10}$ $M_{\odot}/yr$ due to flare like events (Gershberg 2005), while it is only $2\times10^{-14}$ $M_{\odot}/yr$ for the Sun. This suggests that also the angular momentum loss rate is the highest at the early stages of star evolution (also in the pre-main sequence stage according to Marcy and Chen 1992).

On the other hand the mechanism of heating of the stellar corona by the flare activity and causing mass loss is not fully understood. The highest energy detected from two-ribbon flares, which are the most powerful flares occurring on the Sun, is about $10^{30}$ - $10^{31}$ erg (Gershberg 2005, Benz 2008). This level is generally observed in the case of RS CVn type active binaries too (Haisch et al. 1991). However, the observed flare energies vary from $10^{28}$ erg to $10^{34}$ erg in the case of UV Cet type stars of spectral type dMe (Haisch et al. 1991, Gershberg 2005). In addition, some stars of young clusters such as the Pleiades cluster and Orion association exhibit some powerful flare events with $10^{36}$ erg (Gershberg and Shakhovskaya 1983).

There are some remarkable differences between the different types of stars, such as between the Sun itself and UV Cet relevant to the flare energy and the mass loss rate. Nonetheless, most of flare events occurring on a star of the spectral type dMe can be explained by the classical theory of solar flares, in which the main energy source is magnetic reconnection (Gershberg 2005, Hudson and Khan 1996). However, the classical theory of solar flare cannot explain all the flares detected from the UV Cet type stars, which is its shortcoming. Comparison of flare events occurring on different types of stars may improve the observational basis of the theoretical models.

For this purpose, we analyzed the light variations caused by both the rotational modulation due to the stellar cool spots and also the flare events detected in the observation of KIC 7885570, which is an eclipsing binary. Being a binary, KIC 7885570 has some different status as compared to a single UV Cet star. Analyzing the light curve of the system, we find the physical parameters of the components and orbital elements. Then, we analyze the out-of-eclipse variations to find the parameters of the chromospheric activity patterns.

KIC 7885570 (V=11.68 mag) is classified as an eclipsing binary of Algol type. There is no complete light curve or chromospheric activity data for the system listed in the Tycho Input Catalogue by Egret et al. (1992). The 2MASS All-Sky Survey Catalog gives J=10.461 mag, H=10.140 mag, K=10.055 mag (Cutri et al. 2003) for the system listed as 2MASS J19195369+4339137. The binary was observed during the Kepler Mission for a long time with high time resolution (Borucki et al. 2010, Koch et al. 2010, Caldwell et al. 2010). There are several estimates of the temperature of the system and its components. For the ratio of the components radii equal to 3.41 and system inclination $i=74^\circ$.44, Coughlin et al. (2011) found the temperature of the system equal to 5398 K. From the initial analysis of data taken during the Kepler Mission, Slawson et al. (2011) and Pr\v{s}a et al. (2011) found the color excess E(B-V)=0.034 mag, and the temperature ratio of 0.763. Using the all available data, Pinsonneault et al. (2012) found metallicity $[Fe/H]=-0.42$ dex, and the temperature of the primary in the range 5633-5682 K. Huber et al. (2014) derived the temperature of the system in the range 5590-5587 K. Their computed mass and radius of the primary are: $M_{1}=0.819$ $M_{\odot}$ and $R_{1}=0.756$ $R_{\odot}$. Using the data taken by both the 2MASS All-Sky Survey and the Kepler Mission, Armstrong et al. (2014) obtained the temperatures of the components equal to 8254 K and 8233 K, respectively. The orbital period of the system is 1.729348 d (Watson 2006).

There are only a few studies in which the out-of-eclipse variations are analyzed, e.g., Pigulski et al. (2009). The chromospheric activity was studied by Debosscher et al. (2011), and there is only one study about the flare activity observed in the system (Balona 2015).

\section{Data and Analyses}

The Kepler Mission has monitored more than 150 000 targets (Borucki et al. 2010, Koch et al. 2010, Caldwell et al. 2010). The quality of these observations is the highest ever reached in photometry (Jenkins et al. 2010ab). The search for extra-solar planets was the main purpose of the Kepler Mission, but many variable stars such as new eclipsing binaries or pulsating stars have also been discovered (Slawson et al. 2011, Matijevi\v{c} et al. 2012). Some of them exhibit chromospheric activity (Balona 2015).

\begin{figure}[htb]
\includegraphics[width=1.00\textwidth]{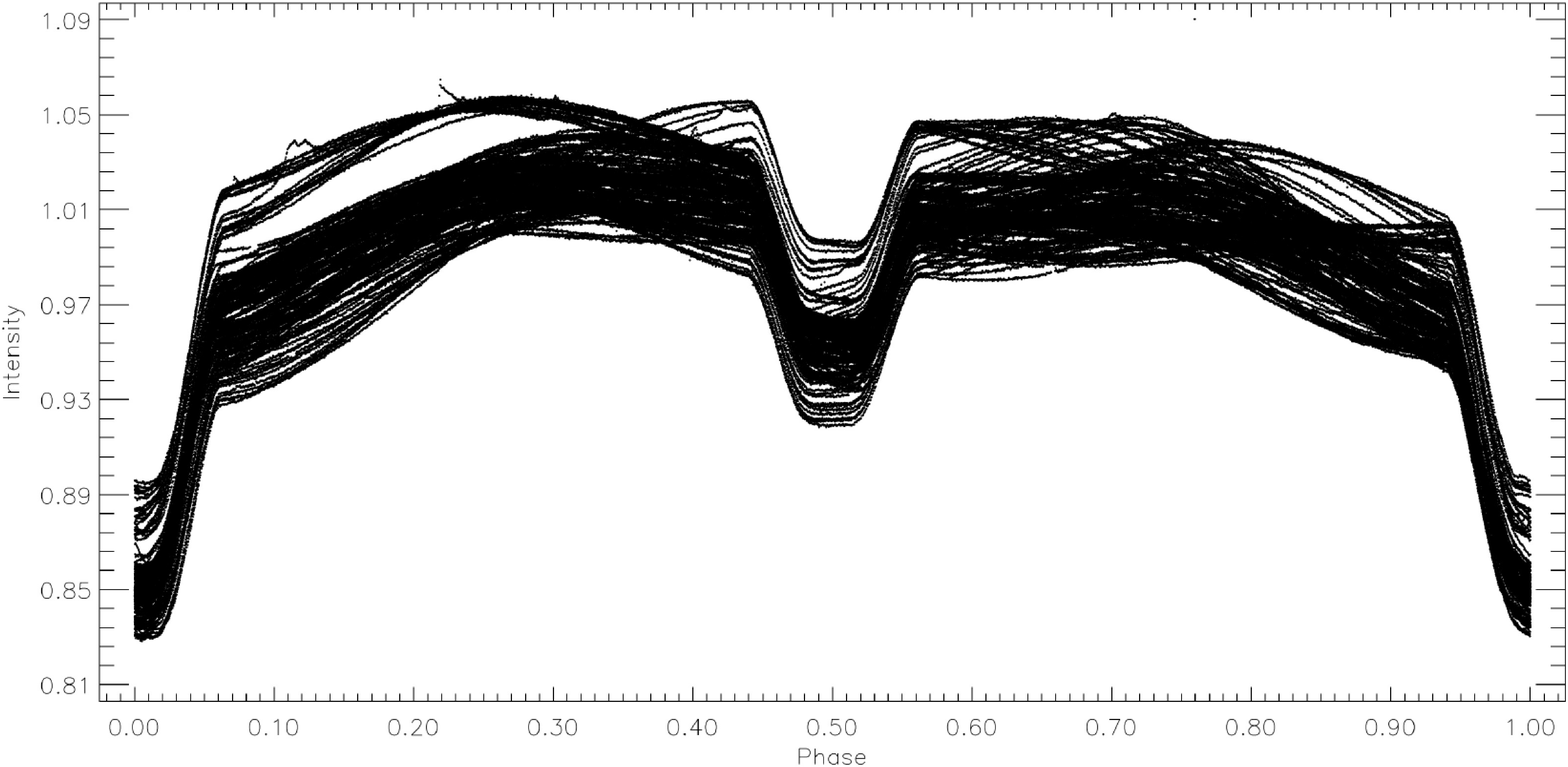}
\FigCap{All the light curves of KIC 7885570 obtained from the available short cadence data in the Kepler Mission Database.}
\label{fig:1}
\end{figure}

In this study, we use the detrended short cadence data to analyze flare activity. The data used in this study were obtained during the Kepler Mission (Slawson et al. 2011, Matijevi\v{c} et al. 2012). The phases are computed using the ephemeris taken from the Kepler Mission database. The resulting light curves are shown in Fig. 1. As can be seen in Fig. 1, there are three dominant variations of the light curve: the eclipses, the sinusoidal variation due to rotational modulation, and the short term flare events. After determining the physical parameters of the components with the light curve analysis, we analyze both the flare and the sinusoidal variations. In the last step, we analyze period variation of the system, using the (O-C) data obtained from the epochs of minima.

\subsection{Light Curve Analysis}

KIC 7885570 was observed during the Kepler Mission for $\approx1300$ d covering over 750 orbital cycles. It can be seen (Fig. 1) that beside eclipses there are two other out-of-eclipse variations, when the light curves are examined cycle by cycle. One of them is an instant-short term variation, while the other one is a strong sinusoidal variation, which varies from one cycle to another. Because of this, each light curve was examined to find the importance of these two variations. We found that there is little sinusoidal light variation effect and no flare variation in the light curve obtained from the data acquired between HJD 2455691.85632 and 2455693.58508. Therefore, the binary system modelling was done with these detrended short cadence data, using the PHOEBE V.0.32 software (Pr\v{s}a and Zwitter 2005), which employs in the 2003 version of the Wilson Devinney Code (Wilson and Devinney 1971, Wilson 1990).

We have tried to analyze the light curve under the following assumptions regarding the evolutionary status of the binary: detached system mode (Mod2), semidetached system with the primary component filling its Roche-Lobe mode (Mod4), and semi-detached system with the secondary component filling its Roche-Lobe mode (Mod5). Our attempts showed that the system is detached, while both semidetached models were rejected.

\begin{figure}[htb]
\includegraphics[width=1.05\textwidth]{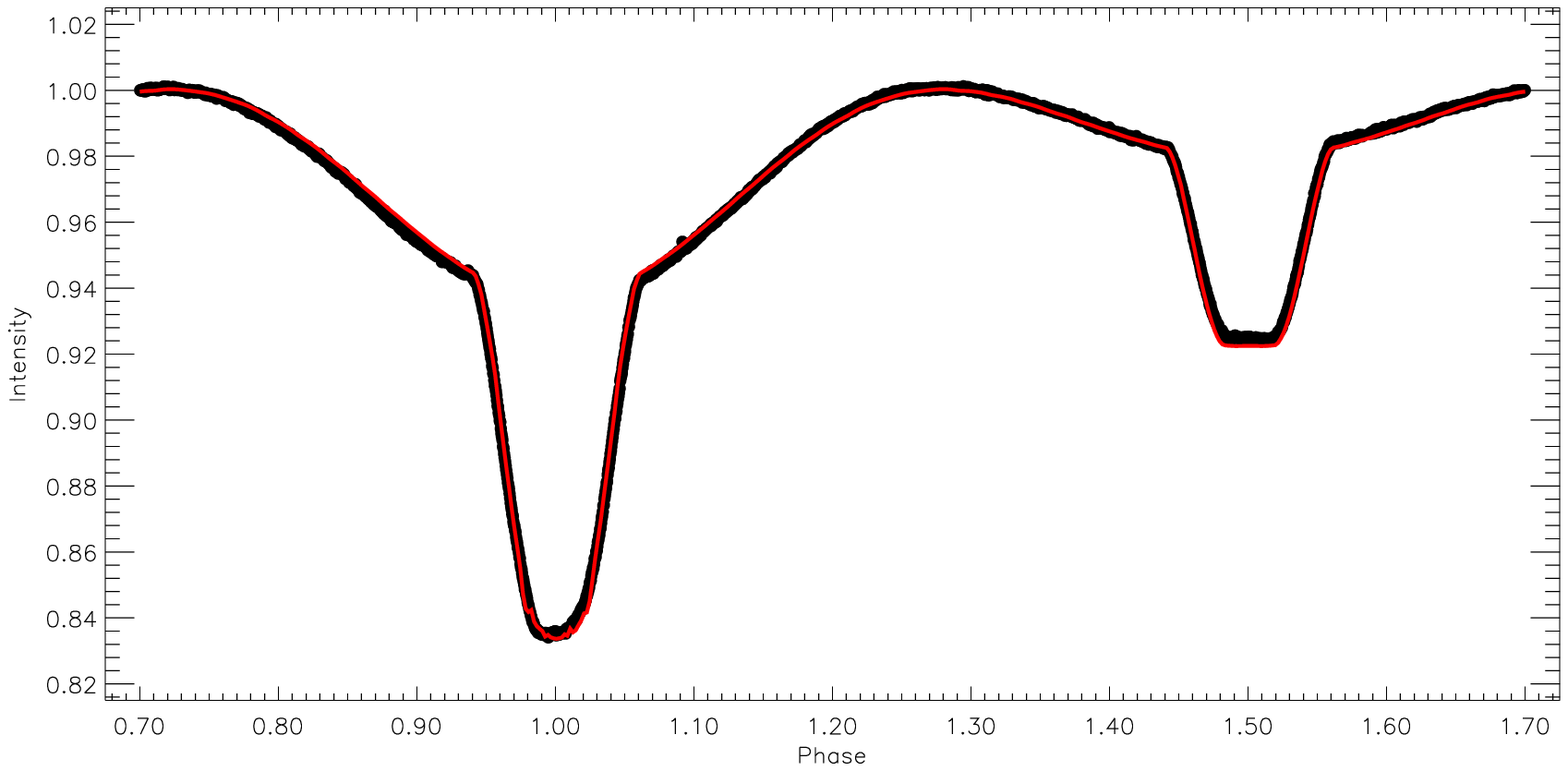}
\vspace{0.1 cm}
\FigCap{The observed (filled circles) and synthetic (red smooth line) light curves obtained from the averaged short cadence data taken from JD 24 55691.85632 and 24 55693.58508.}
\label{fig:2}
\end{figure}

We determine the temperature of the primary component from the $JHK$ photometry of the system, using the values $J=10^{m}.461$, $H=10^{m}.140$, $K=10^{m}.055$ listed by Cutri et al. (2003) in the 2MASS All-Sky Survey Catalog. The derived de-reddened colors ($H-K)_{0}=0^{m}.04$ and ($J-H)_{0}=0^{m}.23$ give the temperature 6530 K with the calibrations of Tokunaga (2000). Thus, we fix the temperature of the primary at 6530 K, while the secondary temperature remains an adjustable parameter.

\begin{table*}
\begin{footnotesize}
\begin{center}
\caption{The physical parameters of components obtained from the Kepler light curve analysis.}
\begin{tabular}{@{}lr@{}}
\hline
Parameter	&	Value	\\
\hline			
$q$	&	0.434$\pm$0.001	\\
$i$ ($^\circ$)	&	80.56$\pm$0.01	\\
$T_{1} (K)$	&	6530 (Fixed)	\\
$T_{2} (K)$	&	5732$\pm$4	\\
$\Omega_{1}$	&	3.7970$\pm$0.0018	\\
$\Omega_{2}$	&	5.7065$\pm$0.0040	\\
$L_{1}/L_{T}$	&	0.9465$\pm$0.0015	\\
$g_{1}, g_{2}$	&	0.32, 0.32 (Fixed)	\\
$A_{1}, A_{2}$	&	0.50, 0.50 (Fixed)	\\
$x_{1},_{bol}, x_{2}, _{bol}$	&	0.680, 0.696 (Fixed)	\\
$x_{1}, x_{2}$	&	0.636, 0.671 (Fixed)	\\
$< r_{1} >$	&	0.3032$\pm$0.0002	\\
$< r_{2} >$	&	0.0983$\pm$0.0001	\\
$Co-Lat_{Spot I}$ $(rad)$	&	1.309$\pm$0.002	\\
$Long_{Spot I}$ $(rad)$	&	3.1416$\pm$0.0001	\\
$R_{Spot I}$ $(rad)$	&	0.995$\pm$0.0002	\\
$T_{f Spot I}$	&	0.65$\pm$0.01	\\
\hline
\end{tabular}
\label{tab1}
\end{center}
\end{footnotesize}
\end{table*}

We use fixed values for the albedos ($A_{1}$ and $A_{2}$) and the gravity darkening coefficients ($g_{1}$ and $g_{2}$) of the components as for stars with the convective envelopes (Lucy 1967, Rucinski 1969). Also the non-linear limb-darkening coefficients ($x_{1}$ and $x_{2}$) are fixed (van Hamme 1993). The other parameters (the dimensionless potentials $\Omega_{1}$ and $\Omega_{2}$, the fractional luminosity $L_{1}$ of the primary, the inclination i of the system, the mass ratio $q$, and the semi-major axis $a$) are all treated as the adjustable free parameters. The best fit (minimum sum of weighted squared residuals $\Sigma res^{2}$ =0.00697) is found for the mass ratio value of $q=0.434$.

Although the cycles which have been chosen for analysis, are little affected by the out-of-eclipse variations, the sinusoidal variation due to the rotational modulation was still seen in the analyzed light curve. Because of this, the out-of-eclipse variation was modeled with the stellar cool spots located on the secondary component.

We assumed that the secondary component was a spotted star for two reasons. The rapid variations in spot positions and sizes in our models (Section 2.3), which is generally seen in the spectral types K and M in the case of the young main sequence stars, point toward the cooler component of the system. In addition, the observed flare activity, characteristic for the stars of the spectral type M can be related to the secondary component.

\subsection{Orbital Period Variation}

The times of minima were computed without any extra corrections on the available short cadence detrended data of the system from the Kepler Mission Database (Slawson et al. 2011, Matijevi\v{c} et al. 2012). For all times of minima, the differences between observations and calculations were computed to determine the residuals $(O-C)_{I}$, using the ephemeris from the Kepler's database:

\begin{center}
\begin{equation}
JD~(Hel.)~=~24~54955.15844~+~1^{d}.7293366~\times~E.
\end{equation}
\end{center}

Some times of minima have very large error. In such cases we examined the light curves to find the sources of errors. It was generally noticed that there is a flare activity in these minima. Therefore, these times were removed from analysis.
Finally, 278 times of minima were obtained from the observations during the Kepler Mission. Using the regression calculations, a linear correction was applied to the differences, and the $(O-C)_{II}$ residuals were obtained. After the linear correction on $(O-C)_{I}$, a new ephemerides was calculated:

\begin{center}
\begin{equation}
JD~(Hel.)~=~24~54955.16445(4)~+~1^{d}.7293365(1)~\times~E.
\end{equation}
\end{center}

\setcounter{table}{1}
\begin{table*}
\begin{footnotesize}
\begin{center}
\caption{Minima times and $(O-C)_{I}$ and $(O-C)_{II}$ residuals.}
\begin{tabular}{@{}cccccccccc@{}}
\hline
HJD	&	$E$	&	Type	&	$(O-C)_{I}$	&	$(O-C)_{II}$	&	HJD	&	$E$	&	Type	&	$(O-C)_{I}$	&	$(O-C)_{II}$	\\
(+24 00000)	&		&		&	(day)	&	day)	&	(+24 00000)	&		&		&	(day)	&	(day)	\\
\hline																			
55002.74831	&	27.5	&	 II 	&	0.03312	&	0.02709	&	55709.15823	&	436.0	&	 I 	&	0.00904	&	0.00306	\\
55003.58186	&	28.0	&	 I 	&	0.00199	&	-0.00403	&	55710.00022	&	436.5	&	 II 	&	-0.01364	&	-0.01962	\\
55004.47610	&	28.5	&	 II 	&	0.03157	&	0.02554	&	55710.88801	&	437.0	&	 I 	&	0.00947	&	0.00350	\\
55005.31987	&	29.0	&	 I 	&	0.01067	&	0.00464	&	55711.72532	&	437.5	&	 II 	&	-0.01788	&	-0.02385	\\
55006.20539	&	29.5	&	 II 	&	0.03153	&	0.02550	&	55712.61874	&	438.0	&	 I 	&	0.01087	&	0.00490	\\
55007.05237	&	30.0	&	 I 	&	0.01383	&	0.00781	&	55713.45374	&	438.5	&	 II 	&	-0.01880	&	-0.02477	\\
55007.93612	&	30.5	&	 II 	&	0.03292	&	0.02689	&	55714.34884	&	439.0	&	 I 	&	0.01164	&	0.00566	\\
55008.78460	&	31.0	&	 I 	&	0.01673	&	0.01071	&	55715.18194	&	439.5	&	 II 	&	-0.01994	&	-0.02591	\\
55009.66647	&	31.5	&	 II 	&	0.03393	&	0.02791	&	55716.07582	&	440.0	&	 I 	&	0.00928	&	0.00330	\\
55010.50355	&	32.0	&	 I 	&	0.00634	&	0.00031	&	55716.90651	&	440.5	&	 II 	&	-0.02470	&	-0.03067	\\
55011.39547	&	32.5	&	 II 	&	0.03360	&	0.02757	&	55717.80126	&	441.0	&	 I 	&	0.00539	&	-0.00059	\\
55012.22844	&	33.0	&	 I 	&	0.00190	&	-0.00413	&	55718.63570	&	441.5	&	 II 	&	-0.02485	&	-0.03082	\\
55013.12529	&	33.5	&	 II 	&	0.03408	&	0.02805	&	55719.52550	&	442.0	&	 I 	&	0.00028	&	-0.00569	\\
55013.95732	&	34.0	&	 I 	&	0.00144	&	-0.00459	&	55720.36601	&	442.5	&	 II 	&	-0.02387	&	-0.02985	\\
55017.41845	&	36.0	&	 I 	&	0.00390	&	-0.00213	&	55721.25173	&	443.0	&	 I 	&	-0.00282	&	-0.00880	\\
55018.31187	&	36.5	&	 II 	&	0.03264	&	0.02662	&	55722.09823	&	443.5	&	 II 	&	-0.02099	&	-0.02697	\\
55019.14443	&	37.0	&	 I 	&	0.00053	&	-0.00549	&	55722.98395	&	444.0	&	 I 	&	0.00006	&	-0.00591	\\
55020.04161	&	37.5	&	 II 	&	0.03306	&	0.02703	&	55723.82738	&	444.5	&	 II 	&	-0.02118	&	-0.02715	\\
55020.87099	&	38.0	&	 I 	&	-0.00223	&	-0.00826	&	55724.72649	&	445.0	&	 I 	&	0.01326	&	0.00729	\\
55021.77136	&	38.5	&	 II 	&	0.03347	&	0.02744	&	55725.55810	&	445.5	&	 II 	&	-0.01980	&	-0.02577	\\
55022.59450	&	39.0	&	 I 	&	-0.00806	&	-0.01409	&	55726.45704	&	446.0	&	 I 	&	0.01448	&	0.00850	\\
55023.49980	&	39.5	&	 II 	&	0.03256	&	0.02654	&	55727.31526	&	446.5	&	 II 	&	0.00803	&	0.00205	\\
55024.31873	&	40.0	&	 I 	&	-0.01317	&	-0.01920	&	55728.18804	&	447.0	&	 I 	&	0.01614	&	0.01017	\\
55025.23112	&	40.5	&	 II 	&	0.03455	&	0.02853	&	55729.04112	&	447.5	&	 II 	&	0.00455	&	-0.00142	\\
55026.04988	&	41.0	&	 I 	&	-0.01136	&	-0.01738	&	55729.89719	&	448.0	&	 I 	&	-0.00405	&	-0.01002	\\
55026.95824	&	41.5	&	 II 	&	0.03234	&	0.02631	&	55730.77515	&	448.5	&	 II 	&	0.00925	&	0.00327	\\
55027.78314	&	42.0	&	 I 	&	-0.00743	&	-0.01346	&	55731.64613	&	449.0	&	 I 	&	0.01556	&	0.00958	\\
55028.68842	&	42.5	&	 II 	&	0.03318	&	0.02716	&	55732.50781	&	449.5	&	 II 	&	0.01258	&	0.00660	\\
55029.51503	&	43.0	&	 I 	&	-0.00488	&	-0.01090	&	55733.37569	&	450.0	&	 I 	&	0.01578	&	0.00980	\\
55030.41801	&	43.5	&	 II 	&	0.03343	&	0.02740	&	55734.23993	&	450.5	&	 II 	&	0.01535	&	0.00938	\\
55031.24825	&	44.0	&	 I 	&	-0.00100	&	-0.00702	&	55735.10168	&	451.0	&	 I 	&	0.01244	&	0.00647	\\
55032.14242	&	44.5	&	 II 	&	0.02850	&	0.02248	&	55735.97329	&	451.5	&	 II 	&	0.01938	&	0.01340	\\
55032.98297	&	45.0	&	 I 	&	0.00438	&	-0.00164	&	55736.82920	&	452.0	&	 I 	&	0.01062	&	0.00465	\\
55156.65457	&	116.5	&	 II 	&	0.02842	&	0.02240	&	55737.70137	&	452.5	&	 II 	&	0.01812	&	0.01215	\\
55157.49406	&	117.0	&	 I 	&	0.00324	&	-0.00278	&	55738.55748	&	453.0	&	 I 	&	0.00956	&	0.00359	\\
55158.38133	&	117.5	&	 II 	&	0.02584	&	0.01982	&	56207.20342	&	724.0	&	 I 	&	0.00528	&	-0.00066	\\
55159.22536	&	118.0	&	 I 	&	0.00521	&	-0.00081	&	56208.06640	&	724.5	&	 II 	&	0.00360	&	-0.00234	\\
55160.10822	&	118.5	&	 II 	&	0.02340	&	0.01738	&	56208.93428	&	725.0	&	 I 	&	0.00681	&	0.00087	\\
55160.95645	&	119.0	&	 I 	&	0.00696	&	0.00094	&	56209.79395	&	725.5	&	 II 	&	0.00181	&	-0.00413	\\
55161.83169	&	119.5	&	 II 	&	0.01752	&	0.01151	&	56210.66441	&	726.0	&	 I 	&	0.00760	&	0.00166	\\
55162.68856	&	120.0	&	 I 	&	0.00973	&	0.00371	&	56211.52180	&	726.5	&	 II 	&	0.00033	&	-0.00561	\\
55163.57005	&	120.5	&	 II 	&	0.02655	&	0.02053	&	56212.39378	&	727.0	&	 I 	&	0.00763	&	0.00169	\\
55164.42028	&	121.0	&	 I 	&	0.01211	&	0.00610	&	56213.25015	&	727.5	&	 II 	&	-0.00066	&	-0.00660	\\
55165.29540	&	121.5	&	 II 	&	0.02257	&	0.01655	&	56214.12327	&	728.0	&	 I 	&	0.00779	&	0.00185	\\
55166.15248	&	122.0	&	 I 	&	0.01498	&	0.00896	&	56214.97698	&	728.5	&	 II 	&	-0.00317	&	-0.00911	\\
55167.02427	&	122.5	&	 II 	&	0.02210	&	0.01608	&	56215.85460	&	729.0	&	 I 	&	0.00978	&	0.00384	\\
55167.88375	&	123.0	&	 I 	&	0.01691	&	0.01089	&	56216.71199	&	729.5	&	 II 	&	0.00250	&	-0.00344	\\
55168.73920	&	123.5	&	 II 	&	0.00769	&	0.00168	&	56217.58431	&	730.0	&	 I 	&	0.01016	&	0.00422	\\
55169.61796	&	124.0	&	 I 	&	0.02179	&	0.01577	&	56218.42239	&	730.5	&	 II 	&	-0.01644	&	-0.02238	\\
55170.47215	&	124.5	&	 II 	&	0.01131	&	0.00529	&	56219.31582	&	731.0	&	 I 	&	0.01233	&	0.00639	\\
55171.34619	&	125.0	&	 I 	&	0.02068	&	0.01467	&	56220.16941	&	731.5	&	 II 	&	0.00125	&	-0.00470	\\
\hline
\end{tabular}
\label{tab2}
\end{center}
\end{footnotesize}
\end{table*}

\setcounter{table}{1}
\begin{table*}
\begin{footnotesize}
\begin{center}
\caption{Continued From Previous Page.}
\begin{tabular}{@{}cccccccccc@{}}
\hline
HJD	&	$E$	&	Type	&	$(O-C)_{I}$	&	$(O-C)_{II}$	&	HJD	&	$E$	&	Type	&	$(O-C)_{I}$	&	$(O-C)_{II}$	\\
(+24 00000)	&		&		&	(day)	&	day)	&	(+24 00000)	&		&		&	(day)	&	(day)	\\
\hline
55172.19796	&	125.5	&	 II 	&	0.00778	&	0.00176	&	56221.04827	&	732.0	&	 I 	&	0.01544	&	0.00951	\\
55173.07251	&	126.0	&	 I 	&	0.01766	&	0.01164	&	56221.87832	&	732.5	&	 II 	&	-0.01918	&	-0.02511	\\
55173.90390	&	126.5	&	 II 	&	-0.01561	&	-0.02163	&	56222.77777	&	733.0	&	 I 	&	0.01561	&	0.00967	\\
55174.78336	&	127.0	&	 I 	&	-0.00083	&	-0.00684	&	56223.60781	&	733.5	&	 II 	&	-0.01902	&	-0.02496	\\
55175.63188	&	127.5	&	 II 	&	-0.01698	&	-0.02299	&	56224.50777	&	734.0	&	 I 	&	0.01627	&	0.01033	\\
55176.50959	&	128.0	&	 I 	&	-0.00394	&	-0.00995	&	56225.36065	&	734.5	&	 II 	&	0.00448	&	-0.00146	\\
55177.36850	&	128.5	&	 II 	&	-0.00969	&	-0.01571	&	56226.24004	&	735.0	&	 I 	&	0.01920	&	0.01326	\\
55178.24244	&	129.0	&	 I 	&	-0.00042	&	-0.00644	&	56227.06474	&	735.5	&	 II 	&	-0.02077	&	-0.02671	\\
55179.09370	&	129.5	&	 II 	&	-0.01382	&	-0.01984	&	56227.96495	&	736.0	&	 I 	&	0.01478	&	0.00884	\\
55179.97302	&	130.0	&	 I 	&	0.00082	&	-0.00519	&	56228.79864	&	736.5	&	 II 	&	-0.01620	&	-0.02214	\\
55180.82075	&	130.5	&	 II 	&	-0.01611	&	-0.02213	&	56229.69068	&	737.0	&	 I 	&	0.01117	&	0.00523	\\
55181.70384	&	131.0	&	 I 	&	0.00231	&	-0.00370	&	56230.54659	&	737.5	&	 II 	&	0.00241	&	-0.00353	\\
55641.70747	&	397.0	&	 I 	&	0.00240	&	-0.00358	&	56231.42025	&	738.0	&	 I 	&	0.01141	&	0.00547	\\
55642.57236	&	397.5	&	 II 	&	0.00263	&	-0.00335	&	56232.27571	&	738.5	&	 II 	&	0.00219	&	-0.00375	\\
55643.43782	&	398.0	&	 I 	&	0.00341	&	-0.00257	&	56233.15045	&	739.0	&	 I 	&	0.01226	&	0.00632	\\
55644.29863	&	398.5	&	 II 	&	-0.00045	&	-0.00643	&	56234.00055	&	739.5	&	 II 	&	-0.00231	&	-0.00824	\\
55645.16605	&	399.0	&	 I 	&	0.00231	&	-0.00367	&	56234.88070	&	740.0	&	 I 	&	0.01318	&	0.00724	\\
55646.02671	&	399.5	&	 II 	&	-0.00170	&	-0.00768	&	56235.73364	&	740.5	&	 II 	&	0.00145	&	-0.00448	\\
55646.89499	&	400.0	&	 I 	&	0.00191	&	-0.00407	&	56236.60967	&	741.0	&	 I 	&	0.01281	&	0.00688	\\
55647.74968	&	400.5	&	 II 	&	-0.00806	&	-0.01404	&	56238.34052	&	742.0	&	 I 	&	0.01432	&	0.00839	\\
55648.62390	&	401.0	&	 I 	&	0.00149	&	-0.00449	&	56239.20686	&	742.5	&	 II 	&	0.01600	&	0.01006	\\
55649.47771	&	401.5	&	 II 	&	-0.00937	&	-0.01535	&	56240.07108	&	743.0	&	 I 	&	0.01555	&	0.00961	\\
55650.35173	&	402.0	&	 I 	&	-0.00003	&	-0.00601	&	56240.94101	&	743.5	&	 II 	&	0.02081	&	0.01487	\\
55651.20608	&	402.5	&	 II 	&	-0.01034	&	-0.01632	&	56241.80224	&	744.0	&	 I 	&	0.01737	&	0.01144	\\
55652.07997	&	403.0	&	 I 	&	-0.00112	&	-0.00710	&	56242.63717	&	744.5	&	 II 	&	-0.01236	&	-0.01830	\\
55652.92512	&	403.5	&	 II 	&	-0.02064	&	-0.02662	&	56243.52947	&	745.0	&	 I 	&	0.01526	&	0.00933	\\
55653.80259	&	404.0	&	 I 	&	-0.00783	&	-0.01381	&	56244.40009	&	745.5	&	 II 	&	0.02122	&	0.01528	\\
55654.65464	&	404.5	&	 II 	&	-0.02045	&	-0.02643	&	56245.25853	&	746.0	&	 I 	&	0.01499	&	0.00906	\\
55655.53157	&	405.0	&	 I 	&	-0.00819	&	-0.01417	&	56252.17055	&	750.0	&	 I 	&	0.00966	&	0.00372	\\
55656.38573	&	405.5	&	 II 	&	-0.01870	&	-0.02468	&	56253.04765	&	750.5	&	 II 	&	0.02209	&	0.01616	\\
55657.26225	&	406.0	&	 I 	&	-0.00685	&	-0.01283	&	56253.89969	&	751.0	&	 I 	&	0.00947	&	0.00353	\\
55658.11684	&	406.5	&	 II 	&	-0.01693	&	-0.02291	&	56254.77715	&	751.5	&	 II 	&	0.02226	&	0.01633	\\
55658.99140	&	407.0	&	 I 	&	-0.00703	&	-0.01301	&	56255.62895	&	752.0	&	 I 	&	0.00939	&	0.00345	\\
55659.84350	&	407.5	&	 II 	&	-0.01960	&	-0.02558	&	56256.50576	&	752.5	&	 II 	&	0.02153	&	0.01560	\\
55660.72121	&	408.0	&	 I 	&	-0.00657	&	-0.01254	&	56257.35942	&	753.0	&	 I 	&	0.01053	&	0.00459	\\
55661.57713	&	408.5	&	 II 	&	-0.01530	&	-0.02128	&	56258.23514	&	753.5	&	 II 	&	0.02157	&	0.01563	\\
55662.45300	&	409.0	&	 I 	&	-0.00411	&	-0.01009	&	56259.08870	&	754.0	&	 I 	&	0.01047	&	0.00453	\\
55663.32383	&	409.5	&	 II 	&	0.00205	&	-0.00393	&	56259.96721	&	754.5	&	 II 	&	0.02430	&	0.01837	\\
55664.18231	&	410.0	&	 I 	&	-0.00413	&	-0.01011	&	56260.81578	&	755.0	&	 I 	&	0.00821	&	0.00228	\\
55665.04958	&	410.5	&	 II 	&	-0.00153	&	-0.00751	&	56261.69435	&	755.5	&	 II 	&	0.02212	&	0.01618	\\
55665.91053	&	411.0	&	 I 	&	-0.00525	&	-0.01123	&	56262.54418	&	756.0	&	 I 	&	0.00728	&	0.00134	\\
55666.78703	&	411.5	&	 II 	&	0.00658	&	0.00060	&	56263.42314	&	756.5	&	 II 	&	0.02157	&	0.01563	\\
55667.64249	&	412.0	&	 I 	&	-0.00263	&	-0.00861	&	56264.27034	&	757.0	&	 I 	&	0.00410	&	-0.00184	\\
55668.51667	&	412.5	&	 II 	&	0.00688	&	0.00090	&	56265.15334	&	757.5	&	 II 	&	0.02243	&	0.01649	\\
55669.36772	&	413.0	&	 I 	&	-0.00673	&	-0.01271	&	56265.99892	&	758.0	&	 I 	&	0.00334	&	-0.00260	\\
55670.25048	&	413.5	&	 II 	&	0.01136	&	0.00538	&	56266.88216	&	758.5	&	 II 	&	0.02191	&	0.01597	\\
55671.09659	&	414.0	&	 I 	&	-0.00720	&	-0.01318	&	56267.72790	&	759.0	&	 I 	&	0.00299	&	-0.00295	\\
55671.98282	&	414.5	&	 II 	&	0.01436	&	0.00838	&	56269.45838	&	760.0	&	 I 	&	0.00413	&	-0.00181	\\
55672.82152	&	415.0	&	 I 	&	-0.01161	&	-0.01759	&	56270.32895	&	760.5	&	 II 	&	0.01003	&	0.00410	\\
55673.70902	&	415.5	&	 II 	&	0.01123	&	0.00525	&	56271.18755	&	761.0	&	 I 	&	0.00396	&	-0.00198	\\
55674.55090	&	416.0	&	 I 	&	-0.01157	&	-0.01755	&	56272.05794	&	761.5	&	 II 	&	0.00968	&	0.00375	\\
\hline
\end{tabular}
\label{tab2}
\end{center}
\end{footnotesize}
\end{table*}

\setcounter{table}{1}
\begin{table*}
\begin{footnotesize}
\begin{center}
\caption{Continued From Previous Page.}
\begin{tabular}{@{}cccccccccc@{}}
\hline
HJD	&	$E$	&	Type	&	$(O-C)_{I}$	&	$(O-C)_{II}$	&	HJD	&	$E$	&	Type	&	$(O-C)_{I}$	&	$(O-C)_{II}$	\\
(+24 00000)	&		&		&	(day)	&	day)	&	(+24 00000)	&		&		&	(day)	&	(day)	\\
\hline
55675.43663	&	416.5	&	 II 	&	0.00950	&	0.00352	&	56272.91770	&	762.0	&	 I 	&	0.00477	&	-0.00116	\\
55676.28279	&	417.0	&	 I 	&	-0.00901	&	-0.01498	&	56273.78605	&	762.5	&	 II 	&	0.00846	&	0.00252	\\
55677.16322	&	417.5	&	 II 	&	0.00675	&	0.00077	&	56274.64810	&	763.0	&	 I 	&	0.00584	&	-0.00010	\\
55678.89492	&	418.5	&	 II 	&	0.00912	&	0.00314	&	56275.51604	&	763.5	&	 II 	&	0.00910	&	0.00317	\\
55679.74916	&	419.0	&	 I 	&	-0.00131	&	-0.00729	&	56276.37654	&	764.0	&	 I 	&	0.00495	&	-0.00099	\\
55680.62492	&	419.5	&	 II 	&	0.00978	&	0.00380	&	56277.24773	&	764.5	&	 II 	&	0.01146	&	0.00552	\\
55681.47994	&	420.0	&	 I 	&	0.00013	&	-0.00584	&	56278.10611	&	765.0	&	 I 	&	0.00518	&	-0.00076	\\
55682.35009	&	420.5	&	 II 	&	0.00561	&	-0.00036	&	56278.97842	&	765.5	&	 II 	&	0.01282	&	0.00688	\\
55683.20932	&	421.0	&	 I 	&	0.00017	&	-0.00581	&	56279.83831	&	766.0	&	 I 	&	0.00804	&	0.00210	\\
55684.08012	&	421.5	&	 II 	&	0.00630	&	0.00033	&	56280.71650	&	766.5	&	 II 	&	0.02156	&	0.01563	\\
55684.94104	&	422.0	&	 I 	&	0.00256	&	-0.00342	&	56281.57083	&	767.0	&	 I 	&	0.01122	&	0.00529	\\
55685.81052	&	422.5	&	 II 	&	0.00737	&	0.00139	&	56282.44634	&	767.5	&	 II 	&	0.02206	&	0.01612	\\
55686.67100	&	423.0	&	 I 	&	0.00319	&	-0.00279	&	56283.30063	&	768.0	&	 I 	&	0.01168	&	0.00574	\\
55687.53104	&	423.5	&	 II 	&	-0.00145	&	-0.00743	&	56284.17625	&	768.5	&	 II 	&	0.02264	&	0.01671	\\
55688.39877	&	424.0	&	 I 	&	0.00162	&	-0.00436	&	56285.02861	&	769.0	&	 I 	&	0.01033	&	0.00439	\\
55689.26303	&	424.5	&	 II 	&	0.00121	&	-0.00477	&	56285.90400	&	769.5	&	 II 	&	0.02104	&	0.01511	\\
55690.12316	&	425.0	&	 I 	&	-0.00333	&	-0.00931	&	56286.75468	&	770.0	&	 I 	&	0.00706	&	0.00113	\\
55690.98790	&	425.5	&	 II 	&	-0.00327	&	-0.00924	&	56287.63191	&	770.5	&	 II 	&	0.01962	&	0.01368	\\
55691.84585	&	426.0	&	 I 	&	-0.00998	&	-0.01595	&	56288.48294	&	771.0	&	 I 	&	0.00599	&	0.00005	\\
55692.71187	&	426.5	&	 II 	&	-0.00862	&	-0.01460	&	56289.36138	&	771.5	&	 II 	&	0.01976	&	0.01382	\\
55693.57218	&	427.0	&	 I 	&	-0.01299	&	-0.01896	&	56290.21326	&	772.0	&	 I 	&	0.00697	&	0.00103	\\
55694.45862	&	427.5	&	 II 	&	0.00879	&	0.00281	&	56291.08996	&	772.5	&	 II 	&	0.01900	&	0.01307	\\
55695.30483	&	428.0	&	 I 	&	-0.00967	&	-0.01565	&	56291.94410	&	773.0	&	 I 	&	0.00847	&	0.00253	\\
55696.16335	&	428.5	&	 II 	&	-0.01582	&	-0.02180	&	56292.82090	&	773.5	&	 II 	&	0.02060	&	0.01467	\\
55697.03614	&	429.0	&	 I 	&	-0.00770	&	-0.01368	&	56293.67592	&	774.0	&	 I 	&	0.01095	&	0.00502	\\
55697.89242	&	429.5	&	 II 	&	-0.01609	&	-0.02207	&	56294.54272	&	774.5	&	 II 	&	0.01309	&	0.00715	\\
55698.76872	&	430.0	&	 I 	&	-0.00445	&	-0.01043	&	56295.40590	&	775.0	&	 I 	&	0.01160	&	0.00566	\\
55699.61994	&	430.5	&	 II 	&	-0.01790	&	-0.02388	&	56296.27452	&	775.5	&	 II 	&	0.01555	&	0.00962	\\
55700.50753	&	431.0	&	 I 	&	0.00501	&	-0.00096	&	56297.13646	&	776.0	&	 I 	&	0.01282	&	0.00689	\\
55701.35795	&	431.5	&	 II 	&	-0.00923	&	-0.01521	&	56298.00600	&	776.5	&	 II 	&	0.01769	&	0.01176	\\
55702.23928	&	432.0	&	 I 	&	0.00743	&	0.00145	&	56298.86704	&	777.0	&	 I 	&	0.01406	&	0.00813	\\
55703.09170	&	432.5	&	 II 	&	-0.00482	&	-0.01079	&	56299.73596	&	777.5	&	 II 	&	0.01831	&	0.01238	\\
55703.97005	&	433.0	&	 I 	&	0.00887	&	0.00289	&	56300.59633	&	778.0	&	 I 	&	0.01402	&	0.00808	\\
55704.82489	&	433.5	&	 II 	&	-0.00096	&	-0.00694	&	56301.46651	&	778.5	&	 II 	&	0.01953	&	0.01359	\\
55705.70050	&	434.0	&	 I 	&	0.00998	&	0.00400	&	56302.32643	&	779.0	&	 I 	&	0.01478	&	0.00885	\\
55706.54829	&	434.5	&	 II 	&	-0.00691	&	-0.01288	&	56303.19366	&	779.5	&	 II 	&	0.01734	&	0.01141	\\
55708.27480	&	435.5	&	 II 	&	-0.00972	&	-0.01570	&	56304.05716	&	780.0	&	 I 	&	0.01617	&	0.01024	\\
\hline
\end{tabular}
\label{tab2}
\end{center}
\end{footnotesize}
\end{table*}

\begin{figure}[htb]
\includegraphics[width=1.10\textwidth]{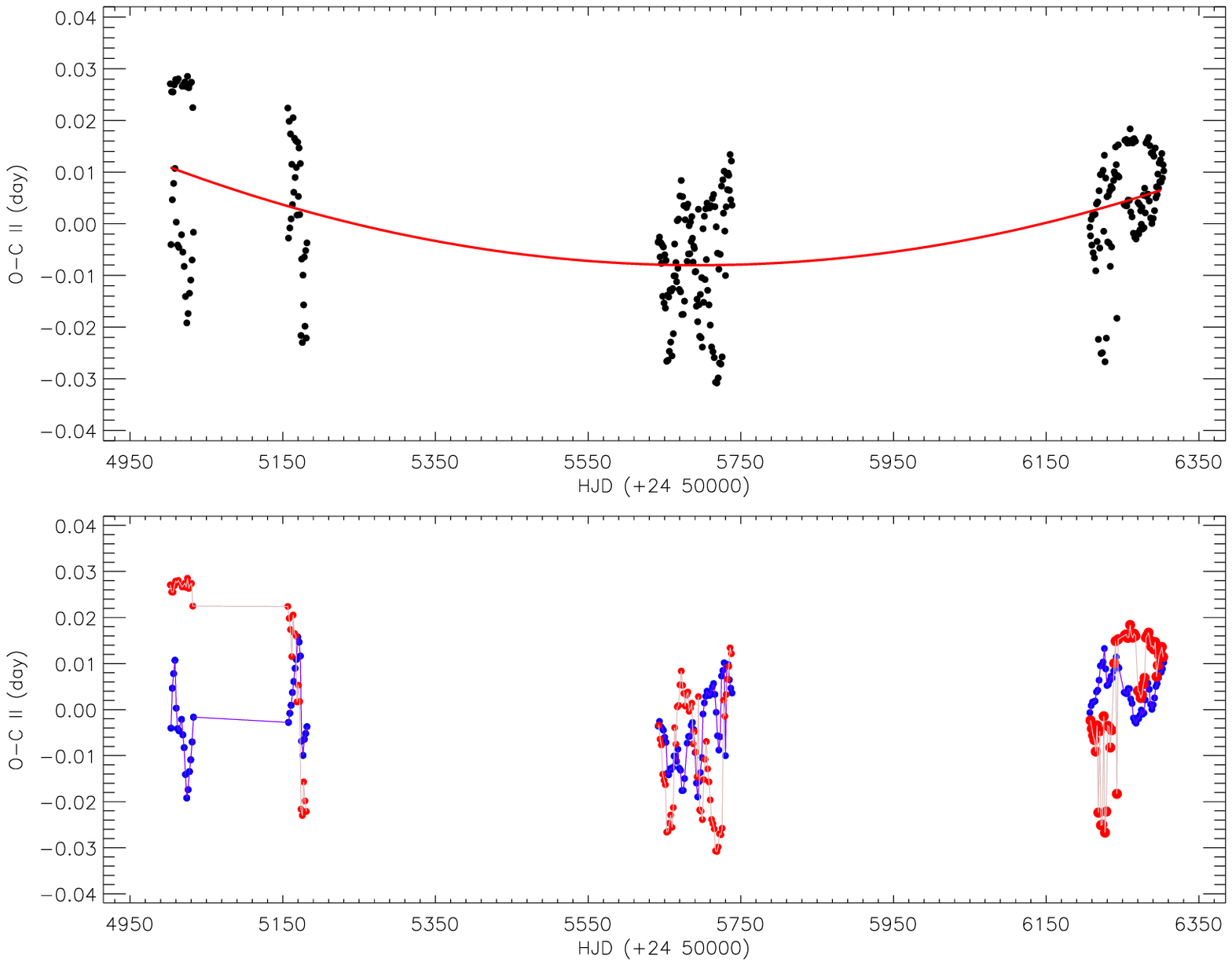}
\vspace{0.2 cm}
\FigCap{ The variation of the $(O-C)_{II}$ residuals obtained by the linear correction on $(O-C)_{I}$. In the upper panel, filled black circles represent the $(O-C)_{II}$ residuals, while smooth red line represents a polynomial fit of the $2^{nd}$ degree derived by the method of least squares. However, in the bottom panel, the $(O-C)_{II}$ residuals are plotted versus time (HJD) for the primary and secondary minima, separately. Here, the filled blue circles represent the primary minima, while the red ones represent the secondary minima.}
\label{fig:3}
\end{figure}

Times of minima, epochs, minimum types, $(O-C)_{I}$ and $(O-C)_{II}$ residuals are listed in Table 2. A sample of the Table 2 is presented here, while full Table 2 is available from the \textit{Acta Astronomica Archive}.

Variations of the $(O-C)_{II}$ residuals are shown in the upper panel of Fig. 3, while the $(O-C)_{II}$ residuals for the primary and secondary minima are plotted in the bottom panel. An interesting phenomenon can be seen in the bottom panel: the residuals of both the primary and the secondary minima vary synchronously, but in opposite directions. A similar phenomenon has been recently demonstrated for other chromospherically active systems by Tran et al. (2013) and Balaji et al. (2015).

\subsection{Rotational Modulation and Stellar Spot Activity}

The sinusoidal variations seen out of eclipses must be caused by the rotational modulation due to the cool stellar spots. After removing other variations from the data (eclipses and flares), the pre-whitened light curves have been obtained. If the sinusoidal variations phased with the orbital period are examined cycle by cycle, it can be seen that both the phases and brightness levels of maxima and minima are rapidly changing in a few cycle time scale. This indicates the rapid evolution of the magnetically active regions. Therefore, the whole pre-whitened light curves cannot be modeled as just one data set. For this reason, the whole data set was divided into several sub-data sets. From the beginning of the data, the variation in the pre-whitened light curves was inspected cycle by cycle. The new sub-data set was started, whenever the difference between the shapes of the two consecutive cycles became bigger than three times the standard deviation. Therefore, the consecutive cycle data, with almost the same phase distributions and brightness levels, were treated as one sub-data set. As a result, the whole data were split into 35 sub-data sets, each of them individually modeled.

\begin{figure}[htb]
\includegraphics[width=1.00\textwidth]{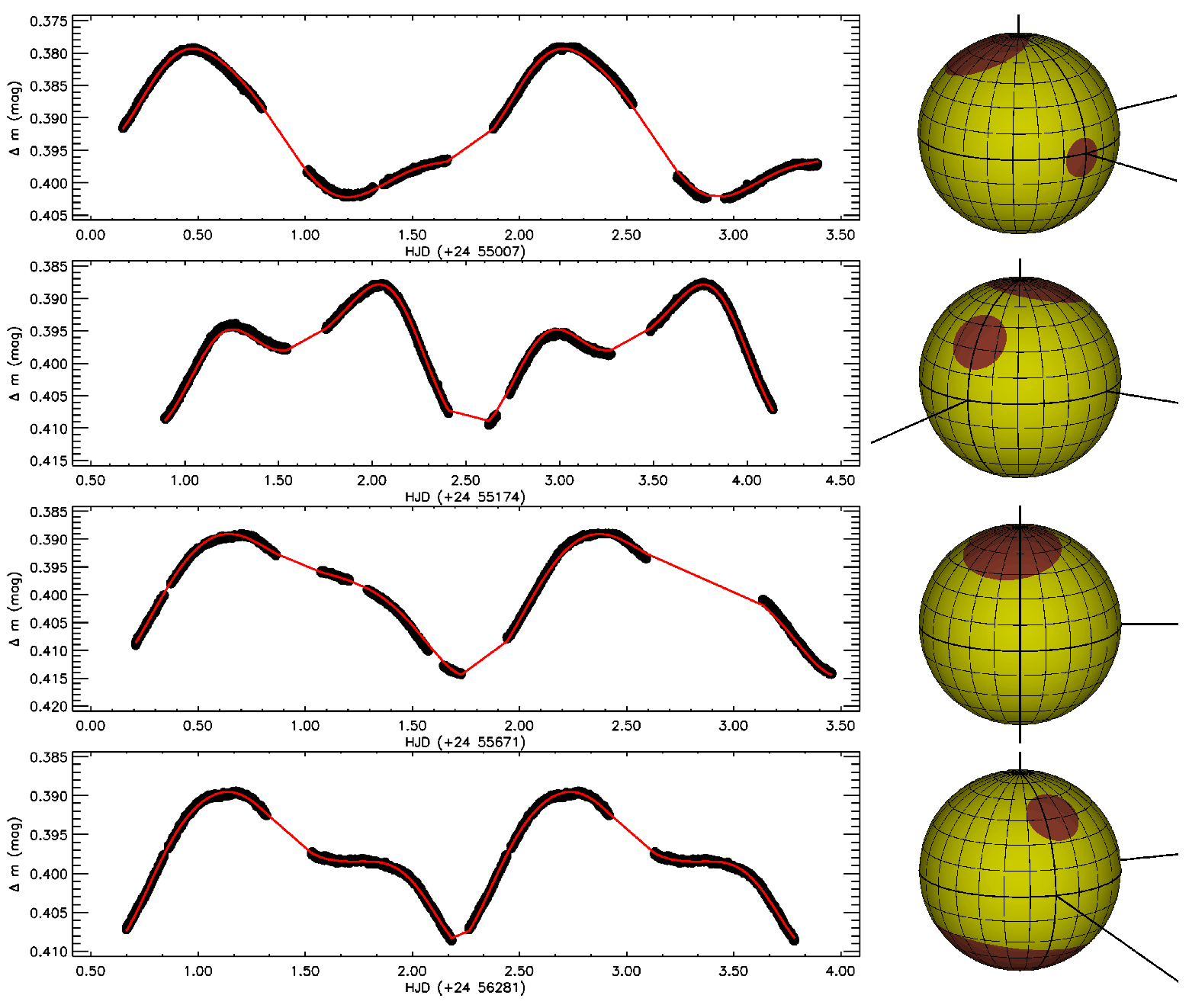}
\vspace{0.1 cm}
\FigCap{ Four examples from 35 models derived for the rotation modulations due to cool spots. In the left panels, the filled circles show the pre-whitened light curve, while the red lines represent the models derived by the SPOTMODEL. In the right panels, the spot distributions on the active component derived by the SPOTMODEL are shown as the 3D form.}
\label{fig:4}
\end{figure}

\setcounter{table}{2}
\begin{table*}
\begin{footnotesize}
\begin{center}
\caption{The parameters of the models derived by the SPOTMODEL program.}
\begin{tabular}{@{}ccccccc@{}}
\hline
Average HJD	&		$b_{1}$		&		$b_{2}$		&		$g_{1}$		&		$g_{2}$		&		$l_{1}$		&		$l_{2}$		\\
(+24 50000)	&		($^\circ$)	&			($^\circ$)	&			($^\circ$)		&		($^\circ$)		&		($^\circ$)		&		($^\circ$)	\\	
\hline																									
5004.03200	&	37.75	$\pm$	0.20	&	117.59	$\pm$	0.72	&	17.49	$\pm$	0.40	&	23.10	$\pm$	0.35	&	-15.71	$\pm$	0.01	&	35.65	$\pm$	0.06	\\
5008.72031	&	12.41	$\pm$	0.09	&	111.25	$\pm$	0.06	&	11.66	$\pm$	0.06	&	26.47	$\pm$	0.03	&	-1.96	$\pm$	0.11	&	60.68	$\pm$	0.14	\\
5012.68289	&	24.19	$\pm$	0.44	&	117.65	$\pm$	0.07	&	12.73	$\pm$	0.02	&	23.94	$\pm$	0.02	&	-2.57	$\pm$	0.08	&	54.01	$\pm$	0.09	\\
5020.06827	&	28.53	$\pm$	0.60	&	118.08	$\pm$	0.15	&	14.61	$\pm$	0.04	&	22.69	$\pm$	0.05	&	-4.41	$\pm$	0.13	&	50.48	$\pm$	0.19	\\
5023.47322	&	29.95	$\pm$	0.54	&	122.29	$\pm$	0.22	&	15.86	$\pm$	0.04	&	21.13	$\pm$	0.06	&	-5.66	$\pm$	0.13	&	47.10	$\pm$	0.20	\\
5026.31184	&	34.42	$\pm$	0.81	&	125.32	$\pm$	0.41	&	15.74	$\pm$	0.07	&	20.76	$\pm$	0.10	&	-6.70	$\pm$	0.22	&	44.41	$\pm$	0.28	\\
5029.93778	&	32.38	$\pm$	0.77	&	122.01	$\pm$	0.28	&	15.64	$\pm$	0.06	&	22.32	$\pm$	0.08	&	-5.92	$\pm$	0.17	&	40.24	$\pm$	0.23	\\
5176.52260	&	35.05	$\pm$	0.41	&	101.44	$\pm$	0.08	&	17.00	$\pm$	0.05	&	28.87	$\pm$	0.04	&	0.51	$\pm$	0.04	&	-20.38	$\pm$	0.19	\\
5180.78777	&	26.57	$\pm$	0.68	&	101.09	$\pm$	0.08	&	15.79	$\pm$	0.05	&	30.08	$\pm$	0.05	&	-0.73	$\pm$	0.07	&	-24.57	$\pm$	0.33	\\
5642.56642	&	64.10	$\pm$	0.29	&	126.12	$\pm$	0.10	&	26.30	$\pm$	0.10	&	16.41	$\pm$	0.23	&	-7.17	$\pm$	0.05	&	-1.78	$\pm$	0.09	\\
5649.31378	&	68.13	$\pm$	0.07	&	162.47	$\pm$	0.09	&	28.42	$\pm$	0.02	&	13.07	$\pm$	0.02	&	-14.58	$\pm$	0.05	&	-14.90	$\pm$	0.09	\\
5655.91200	&	66.63	$\pm$	0.11	&	126.75	$\pm$	0.76	&	29.12	$\pm$	0.05	&	15.14	$\pm$	0.14	&	-19.46	$\pm$	0.04	&	-21.70	$\pm$	0.10	\\
5659.18155	&	69.09	$\pm$	0.07	&	138.37	$\pm$	0.86	&	29.52	$\pm$	0.03	&	13.23	$\pm$	0.09	&	-24.51	$\pm$	0.05	&	-13.65	$\pm$	0.13	\\
5663.14341	&	68.90	$\pm$	0.09	&	131.33	$\pm$	0.75	&	29.45	$\pm$	0.04	&	14.08	$\pm$	0.12	&	-24.29	$\pm$	0.06	&	-12.25	$\pm$	0.10	\\
5668.95455	&	69.11	$\pm$	0.09	&	115.85	$\pm$	0.77	&	29.68	$\pm$	0.60	&	16.15	$\pm$	0.71	&	-20.43	$\pm$	0.41	&	-9.98	$\pm$	0.82	\\
5672.69108	&	70.67	$\pm$	0.80	&	116.29	$\pm$	0.79	&	29.42	$\pm$	0.50	&	17.97	$\pm$	0.24	&	-15.36	$\pm$	0.85	&	11.51	$\pm$	0.20	\\
5676.80771	&	68.02	$\pm$	0.11	&	121.31	$\pm$	0.81	&	28.22	$\pm$	0.53	&	18.75	$\pm$	0.16	&	-11.06	$\pm$	0.62	&	6.48	$\pm$	0.73	\\
5683.96383	&	64.00	$\pm$	0.25	&	118.37	$\pm$	0.83	&	26.66	$\pm$	0.55	&	21.74	$\pm$	0.92	&	-7.21	$\pm$	0.28	&	2.70	$\pm$	0.39	\\
5691.52471	&	45.64	$\pm$	0.67	&	104.96	$\pm$	0.33	&	20.78	$\pm$	0.17	&	20.09	$\pm$	0.15	&	-1.45	$\pm$	0.04	&	3.95	$\pm$	0.13	\\
5709.48021	&	42.08	$\pm$	0.35	&	122.35	$\pm$	0.49	&	23.36	$\pm$	0.06	&	19.34	$\pm$	0.11	&	2.73	$\pm$	0.08	&	-26.06	$\pm$	0.23	\\
5713.52167	&	43.58	$\pm$	0.48	&	114.78	$\pm$	0.56	&	23.53	$\pm$	0.11	&	20.52	$\pm$	0.18	&	-1.53	$\pm$	0.14	&	-41.00	$\pm$	0.57	\\
5717.33918	&	55.79	$\pm$	0.16	&	129.53	$\pm$	0.75	&	26.74	$\pm$	0.06	&	16.70	$\pm$	0.14	&	-0.61	$\pm$	0.14	&	-44.82	$\pm$	0.33	\\
5728.18173	&	69.85	$\pm$	0.02	&	238.41	$\pm$	0.77	&	32.34	$\pm$	0.01	&	10.64	$\pm$	0.35	&	8.08	$\pm$	0.03	&	-42.31	$\pm$	0.07	\\
5731.62378	&	68.73	$\pm$	0.05	&	244.09	$\pm$	0.79	&	31.95	$\pm$	0.02	&	10.10	$\pm$	0.56	&	9.30	$\pm$	0.06	&	-32.14	$\pm$	0.07	\\
6230.51567	&	42.81	$\pm$	0.32	&	263.64	$\pm$	0.09	&	18.78	$\pm$	0.08	&	55.91	$\pm$	0.06	&	7.74	$\pm$	0.03	&	-4.34	$\pm$	0.17	\\
6234.02066	&	37.91	$\pm$	0.71	&	267.06	$\pm$	0.14	&	17.36	$\pm$	0.13	&	57.42	$\pm$	0.09	&	10.36	$\pm$	0.07	&	-2.25	$\pm$	0.68	\\
6241.79821	&	39.10	$\pm$	0.50	&	263.76	$\pm$	0.13	&	19.08	$\pm$	0.11	&	50.20	$\pm$	0.11	&	11.77	$\pm$	0.07	&	59.75	$\pm$	0.17	\\
6255.60165	&	43.95	$\pm$	0.40	&	256.65	$\pm$	0.11	&	20.47	$\pm$	0.11	&	50.28	$\pm$	0.12	&	9.34	$\pm$	0.12	&	67.98	$\pm$	0.60	\\
6262.55028	&	43.82	$\pm$	0.38	&	257.77	$\pm$	0.10	&	19.74	$\pm$	0.09	&	51.03	$\pm$	0.10	&	3.16	$\pm$	0.09	&	54.90	$\pm$	0.49	\\
6266.00699	&	49.31	$\pm$	0.37	&	256.93	$\pm$	0.15	&	20.62	$\pm$	0.12	&	49.19	$\pm$	0.14	&	0.87	$\pm$	0.12	&	48.35	$\pm$	0.58	\\
6269.60959	&	33.21	$\pm$	0.90	&	263.26	$\pm$	0.06	&	15.36	$\pm$	0.12	&	53.27	$\pm$	0.09	&	5.05	$\pm$	0.12	&	76.36	$\pm$	0.21	\\
6273.80010	&	34.16	$\pm$	0.58	&	264.27	$\pm$	0.02	&	14.35	$\pm$	0.07	&	55.23	$\pm$	0.04	&	2.75	$\pm$	0.07	&	86.96	$\pm$	0.73	\\
6282.43076	&	47.03	$\pm$	0.83	&	262.77	$\pm$	0.13	&	16.01	$\pm$	0.20	&	56.30	$\pm$	0.11	&	8.59	$\pm$	0.14	&	54.49	$\pm$	0.95	\\
6286.77790	&	49.15	$\pm$	0.69	&	264.95	$\pm$	0.12	&	16.91	$\pm$	0.20	&	56.35	$\pm$	0.12	&	13.13	$\pm$	0.11	&	59.14	$\pm$	0.38	\\
6292.81678	&	65.78	$\pm$	0.31	&	250.68	$\pm$	0.40	&	27.28	$\pm$	0.15	&	42.52	$\pm$	0.40	&	11.81	$\pm$	0.12	&	46.35	$\pm$	0.22	\\
\hline
\end{tabular}
\label{tab3}
\end{center}
\end{footnotesize}
\end{table*}

Using the SPOTMODEL program (Rib\'{a}rik 2002, Rib\'{a}rik et al. 2003), the subdata sets were modeled under some assumptions to derive the spot distribution on the stellar surface and to find their parameters such as the spot radius, latitude and especially longitude. To model any spot distribution on the stellar surface, the SPOTMODEL program needs two-band observations or spot temperature factor parameter. However, the available data taken from the Kepler Mission Database contain only monochromatic observations. At this point, considering the results obtained from the light curve analysis of the system and also the results found for other similar systems (Clausen et al. 2001, Thomas and Weiss 2008), it was assumed that this is the secondary component which exhibits chromospheric activity. According to the light curve analysis, there are two spotted areas on the secondary component. One spotted area or more than two do not give any acceptable fit to the observations. 

We assumed that the spot temperature factor parameter should be in the range of 0.70-0.95 in the SPOTMODEL. Our tests indicated that the best solutions are obtained if the spot temperature factor was taken as $T_{factor}=0.75$ for the primary spot (Spot 1), and as $T_{factor}=0.80$ for the secondary one (Spot 2). Therefore, it was assumed that the spot temperature factors were constant parameters for all sub-data set, while the other parameters such as the longitudes ($I$), latitudes ($b$) and radii of the spots ($g$), were treated as the adjustable free parameters for each sub-data set separately.

Fig. 4 shows four models from our 35 models of photometric sub-data sets. Model fits and the distribution of cool spot on the 3D plot are presented. The spot parameters derived by SPOTMODEL are listed for all sub-data sets in Table 3. The average of the HJD interval for each sub-data set, spot latitudes ($b$), radii of the spots ($g$) and spot longitudes ($I$) are listed there. As can be seen from Table 3, the spot parameters remain constant for several cycles and then change by several degrees when going from one cycle to another. This is a common phenomenon seen in the case of the chromospherically active young stars like FL Lyr (Yoldas and Dal 2016). It is well known that the chromospheric activity patterns can in general change rapidly in short time intervals (Gershberg 2005, Benz 2008). The variations of spot parameters vs. time are shown in Fig. 5. It should be noted here that if it was assumed that chromospherically active star is not the secondary component, but the primary one, there would be no astrophysically reasonable solution and distinctive changes in the values of spot latitudes ($I$), radii of the spots ($g$) and spot longitudes ($b$). This is because the surface temperatures of the both components are substantially different.

\begin{figure}[htb]
\includegraphics[width=1.05\textwidth]{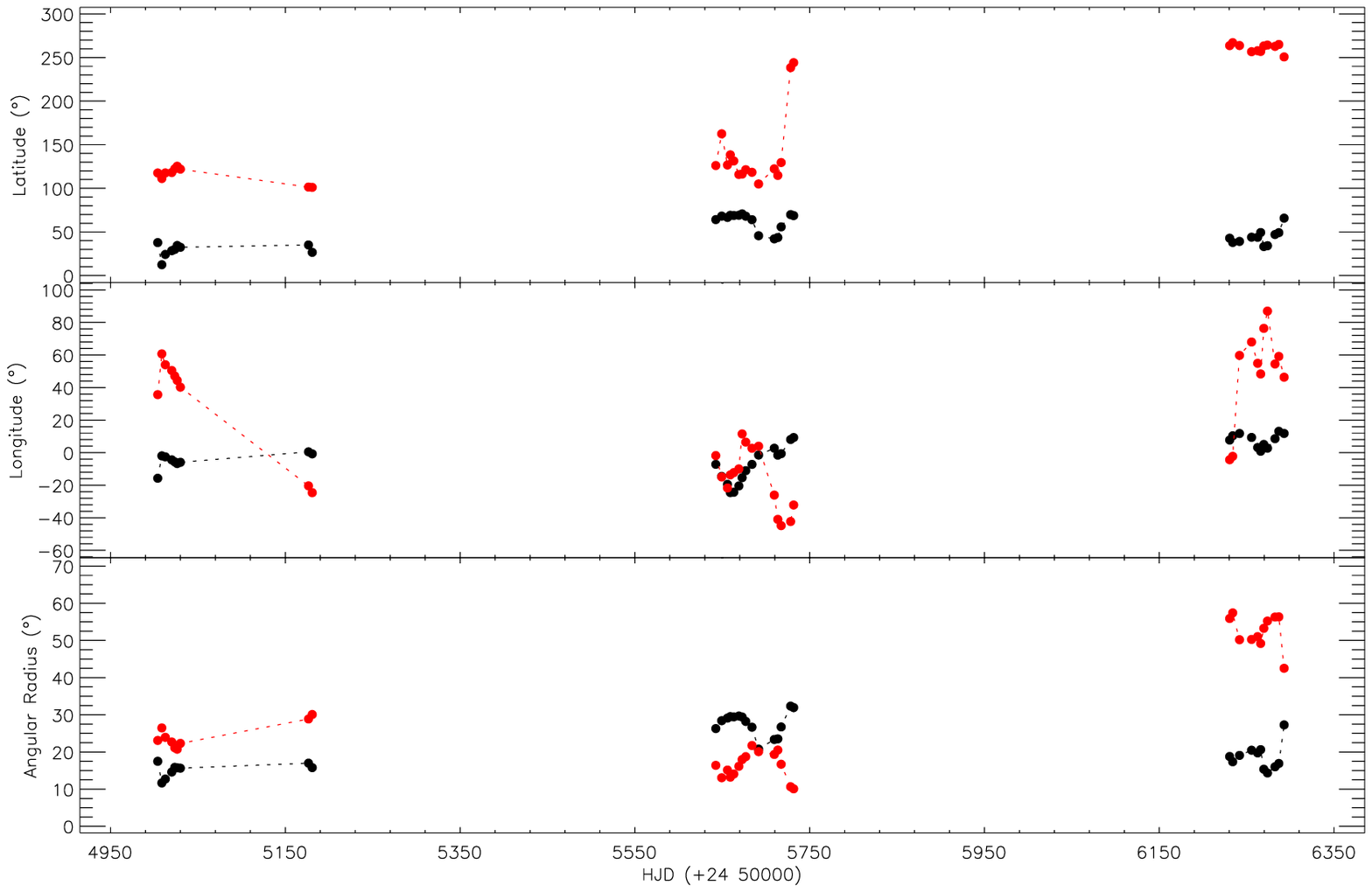}
\FigCap{The variations of the free parameters found by the SPOTMODEL Program. The filled black circles represent the Spot 1 ($T_{factor}=0.75$), while filled red circles represent the Spot 2 ($T_{factor}=0.80$) in the figure.}
\label{fig:5}
\end{figure}

\subsection{Flare Activity and the OPEA Model}

To analyze the flare activity in KIC 7885570, we removed all other variations from light the curve. Observations between the phases 0.96-0.04 and 0.46-0.54 (related to the primary and secondary minima) were neglected. Also all observations with large error due to technical problems were removed from the data sets. As described in Section 2.3 the sinusoidal variation in the light curve has been fitted based on data out of eclipses and flares. The whole dataset can now be compared to the synthetic light curves obtained in this way in a search for short lasting intensity excess, i.e., flares. Three flare light curves and the synthetic light curves are shown in Fig. 6. Comparing the data with the model light curve, the flare rise times ($T_{r}$), decay times ($T_{d}$), the flare amplitudes ($A$) and finally the flare equivalent durations ($P$) were computed. The parameters of flares are listed in Table 4. A sample part of Table 4 is presented here, while full table is available from the \textit{Acta Astronomica Archive}.

\begin{figure}[htb]
\hspace{1.0 cm}
\includegraphics[width=1.60\textwidth]{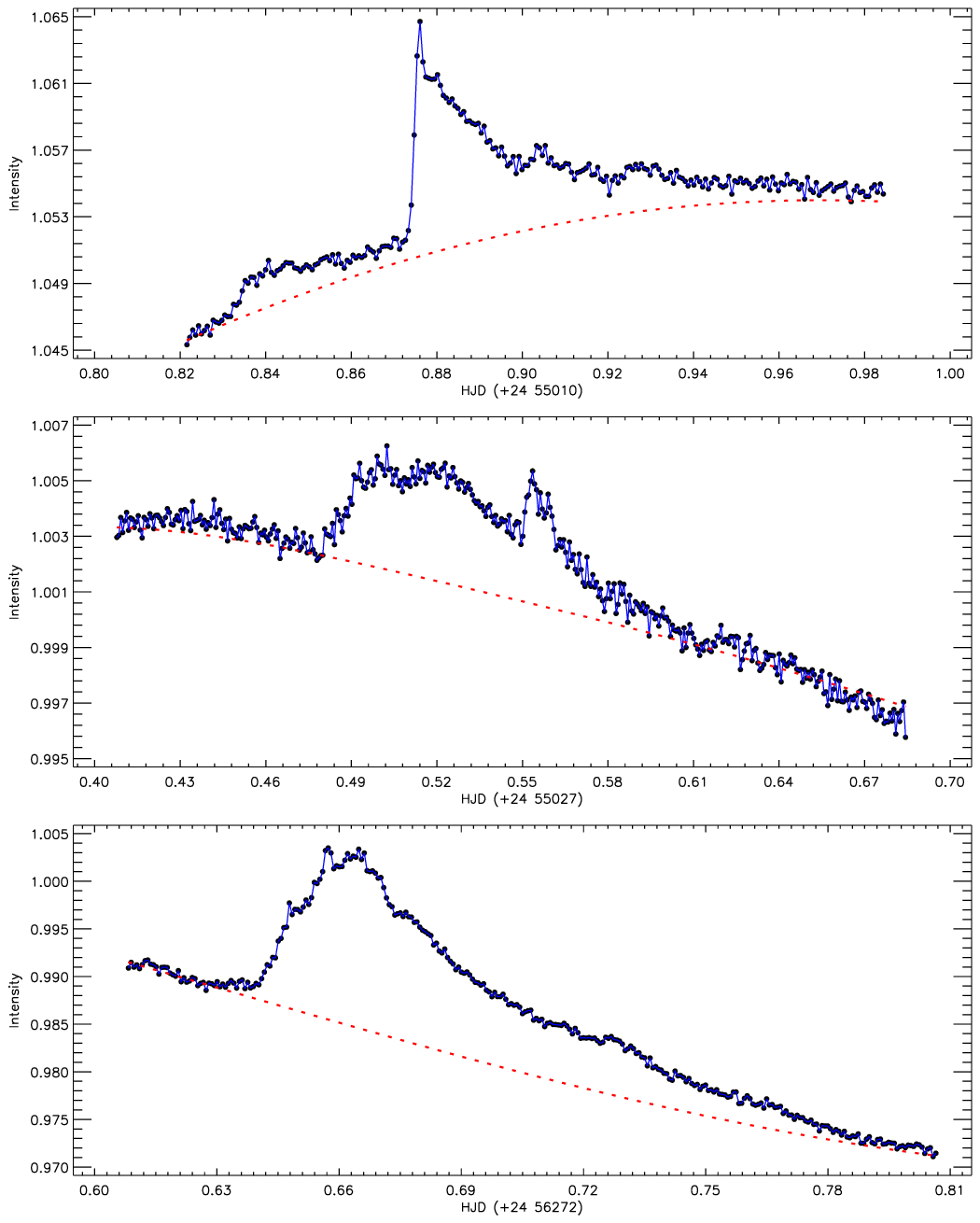}
\vspace{0.3 cm}
\FigCap{There flare light curve samples. In the figure, filled circles represent the observations, while the dotted lines represent the Fourier models.}
\label{fig:6}
\end{figure}

\setcounter{table}{3}
\begin{table*}
\begin{footnotesize}
\begin{center}
\caption{The flare parameters computed from the available Short Cadence Data in the Kepler Mission database.}
\begin{tabular}{@{}crrrc@{}}
\hline
Flare Time	&	$P$	&	$T_{r}$	&	$T_{d}$	&	$Amplitude$	\\
(+24 00000)	&	(s)	&	(s)	&	(s)	&	(Intensity)	\\
\hline
55002.897228	&	10.67513	&	1588.95475	&	5708.47565	&	0.00416	\\
55002.980327	&	8.65908	&	353.09952	&	5119.96378	&	0.00511	\\
55004.100798	&	27.57592	&	1059.30806	&	7297.42608	&	0.01108	\\
55004.626635	&	11.30656	&	529.65360	&	6532.37568	&	0.00309	\\
55008.327249	&	2.85669	&	1235.85091	&	2765.96122	&	0.00185	\\
55010.620632	&	6.48999	&	411.96384	&	2942.49370	&	0.00626	\\
55010.876057	&	32.90377	&	7768.21018	&	3825.25286	&	0.01408	\\
55011.592611	&	7.65157	&	588.50842	&	5296.50230	&	0.00213	\\
55018.732927	&	1.02885	&	235.40976	&	823.88966	&	0.00161	\\
55018.749274	&	1.54401	&	588.49805	&	1471.24426	&	0.00154	\\
55018.771751	&	0.35084	&	470.80138	&	470.79274	&	0.00108	\\
55019.342539	&	97.19761	&	6179.22346	&	9121.71024	&	0.01868	\\
55020.618979	&	2.49765	&	588.49632	&	1294.69018	&	0.00271	\\
55021.104625	&	3.73807	&	1765.49933	&	1824.33859	&	0.00175	\\
55022.103843	&	2.01406	&	117.69581	&	3119.02618	&	0.00123	\\
55023.766481	&	1.69797	&	529.64755	&	2236.27910	&	0.00146	\\
55025.866401	&	13.71493	&	1942.03699	&	2059.72416	&	0.00763	\\
55027.306306	&	1.23595	&	1000.43770	&	1118.14992	&	0.00139	\\
55027.502470	&	26.28887	&	1942.03440	&	9474.75014	&	0.00445	\\
55030.133664	&	0.62521	&	235.39075	&	470.78928	&	0.00237	\\
55030.577078	&	0.61956	&	176.54285	&	647.34941	&	0.00482	\\
55030.586613	&	0.39579	&	176.55149	&	470.78928	&	0.00167	\\
55032.270357	&	1.09919	&	765.04349	&	588.49286	&	0.00141	\\
55032.281936	&	0.74200	&	411.94051	&	823.89139	&	0.00131	\\
55175.292675	&	0.39886	&	176.54371	&	58.83667	&	0.00609	\\
55176.697777	&	4.29154	&	1941.94627	&	1824.25651	&	0.00241	\\
55178.366464	&	0.23733	&	117.69062	&	58.83667	&	0.00343	\\
55179.853981	&	1.58539	&	353.07878	&	823.86634	&	0.00485	\\
55181.170544	&	0.77598	&	176.53507	&	235.38989	&	0.01071	\\
55182.059377	&	8.47162	&	411.94224	&	4236.97478	&	0.00688	\\
55648.070187	&	3.38497	&	823.91040	&	1530.11549	&	0.00292	\\
55649.839794	&	6.08832	&	706.21373	&	4119.54077	&	0.00446	\\
55650.736178	&	18.74595	&	2412.87811	&	7532.88768	&	0.00502	\\
55651.834181	&	128.60604	&	3825.29866	&	11652.46128	&	0.02497	\\
55654.184807	&	14.47501	&	353.11248	&	8121.41510	&	0.00574	\\
55656.132879	&	6.16941	&	823.90522	&	4237.27459	&	0.00387	\\
55656.235732	&	78.13174	&	2824.84454	&	5590.84032	&	0.03034	\\
55656.712533	&	10.75699	&	823.90522	&	4649.22893	&	0.00319	\\
55657.663411	&	6.82572	&	1530.11290	&	4531.51498	&	0.00239	\\
55658.317991	&	4.32212	&	1294.71610	&	2707.14010	&	0.00225	\\
55659.549501	&	9.01695	&	2883.69763	&	2707.14960	&	0.00408	\\
55660.458830	&	19.09031	&	941.61312	&	7768.32941	&	0.00512	\\
55660.559639	&	2.94486	&	588.50928	&	2059.78291	&	0.00393	\\
55661.459432	&	8.92595	&	2295.19786	&	3354.50074	&	0.00284	\\
55662.095622	&	2.64608	&	353.10384	&	1706.67994	&	0.00341	\\
55663.424538	&	0.36284	&	58.84963	&	117.70704	&	0.00545	\\
55664.664224	&	25.26883	&	1059.32102	&	8297.99251	&	0.00573	\\
55668.161911	&	3.95248	&	941.62349	&	3295.65888	&	0.00237	\\
55668.803552	&	2.27129	&	1294.71178	&	1471.28573	&	0.00155	\\
55670.835413	&	2.59642	&	1118.18189	&	1647.83462	&	0.00171	\\
55671.519285	&	0.39148	&	294.25507	&	294.26458	&	0.00248	\\
\hline
\end{tabular}
\label{tab4}
\end{center}
\end{footnotesize}
\end{table*}

\setcounter{table}{3}
\begin{table*}
\begin{footnotesize}
\begin{center}
\caption{Continued From Previous Page.}
\begin{tabular}{@{}crrrc@{}}
\hline
Flare Time	&	$P$	&	$T_{r}$	&	$T_{d}$	&	$Amplitude$	\\
(+24 00000)	&	(s)	&	(s)	&	(s)	&	(Intensity)	\\
\hline
55672.224954	&	2.66442	&	1471.27795	&	1942.09056	&	0.00145	\\
55672.253562	&	0.43699	&	117.70790	&	588.51101	&	0.00121	\\
55672.599585	&	4.95169	&	1294.72128	&	2766.00874	&	0.00255	\\
55674.090617	&	2.90422	&	706.21027	&	2295.19613	&	0.00163	\\
55675.313277	&	5.52057	&	529.66224	&	1647.83549	&	0.00435	\\
55678.627062	&	12.28612	&	411.96384	&	5296.60685	&	0.00824	\\
55679.507787	&	1.02091	&	411.95434	&	823.91818	&	0.00193	\\
55682.625403	&	0.52162	&	235.39853	&	529.66224	&	0.00234	\\
55685.970523	&	1.56884	&	588.51274	&	1647.82858	&	0.00143	\\
55691.100934	&	11.86987	&	529.67174	&	4001.87779	&	0.00798	\\
55693.982193	&	5.79558	&	1588.98845	&	2177.49168	&	0.00272	\\
55694.948061	&	12.58234	&	353.10557	&	3236.81616	&	0.01100	\\
55697.759842	&	7.42440	&	1412.42141	&	2589.46330	&	0.00377	\\
55702.705658	&	146.22706	&	2118.64032	&	14771.65853	&	0.02786	\\
55708.990607	&	1.48384	&	176.54803	&	941.62435	&	0.00235	\\
55712.093912	&	11.57002	&	1530.12586	&	5590.85069	&	0.00426	\\
55714.529692	&	12.60495	&	529.66138	&	3177.96480	&	0.00760	\\
55716.528856	&	0.80430	&	529.66051	&	529.66051	&	0.00145	\\
55717.573052	&	12.20767	&	823.90694	&	5943.95539	&	0.00380	\\
55718.229677	&	1.02910	&	529.66915	&	765.04867	&	0.00137	\\
55718.239213	&	0.29040	&	58.85741	&	588.51014	&	0.00118	\\
55718.382253	&	5.66139	&	1118.16115	&	4119.56755	&	0.00253	\\
55719.820832	&	0.28776	&	117.69840	&	294.25421	&	0.00297	\\
55720.765580	&	0.60883	&	235.39680	&	529.65965	&	0.00319	\\
55727.607674	&	3.16930	&	882.76954	&	1883.22192	&	0.00263	\\
55730.895551	&	0.69733	&	235.40458	&	529.66656	&	0.00197	\\
55731.353278	&	2.09393	&	882.76003	&	882.76867	&	0.00257	\\
55731.761283	&	2.57682	&	1294.71955	&	1588.95648	&	0.00183	\\
55731.780355	&	0.41303	&	58.85741	&	411.95088	&	0.00238	\\
55732.360688	&	5.69604	&	1000.46707	&	2471.73293	&	0.00523	\\
55732.392020	&	0.81147	&	235.39507	&	529.65792	&	0.00190	\\
55736.629401	&	0.35955	&	235.41235	&	176.55494	&	0.00295	\\
56231.120993	&	3.55548	&	765.00634	&	1706.54774	&	0.00338	\\
56231.656332	&	1.99454	&	411.92150	&	1647.70243	&	0.00241	\\
56235.156467	&	0.88557	&	470.76509	&	411.93878	&	0.00159	\\
56240.815658	&	3.00616	&	1647.69120	&	588.47126	&	0.00321	\\
56241.066300	&	12.79948	&	470.77286	&	4707.70272	&	0.00622	\\
56241.562815	&	20.98885	&	647.30707	&	4236.93677	&	0.01036	\\
56242.067504	&	0.84725	&	529.61818	&	764.99683	&	0.00114	\\
56254.097596	&	1.81944	&	411.91978	&	1706.53306	&	0.00196	\\
56255.127405	&	4.31030	&	294.23952	&	3001.15498	&	0.00212	\\
56255.865707	&	4.12307	&	1176.91574	&	2236.15901	&	0.00253	\\
56256.935700	&	8.94963	&	1294.62192	&	5001.92755	&	0.00407	\\
56260.990914	&	4.99894	&	1412.31082	&	2648.08915	&	0.00453	\\
56263.640357	&	6.43391	&	706.16016	&	4472.32147	&	0.00398	\\
56265.358068	&	1.06821	&	235.37779	&	941.54659	&	0.00146	\\
56265.415961	&	41.10177	&	4060.39392	&	7767.70128	&	0.00994	\\
56266.119528	&	21.00597	&	1176.93302	&	6473.09750	&	0.00686	\\
56267.019250	&	2.38824	&	706.14288	&	2236.16160	&	0.00196	\\
56268.021135	&	1.38561	&	294.22310	&	1353.46810	&	0.00186	\\
56270.590893	&	7.33015	&	706.16102	&	4178.08627	&	0.00404	\\
\hline
\end{tabular}
\label{tab4}
\end{center}
\end{footnotesize}
\end{table*}

\setcounter{table}{3}
\begin{table*}
\begin{footnotesize}
\begin{center}
\caption{Continued From Previous Page.}
\begin{tabular}{@{}crrrc@{}}
\hline
Flare Time	&	$P$	&	$T_{r}$	&	$T_{d}$	&	$Amplitude$	\\
(+24 00000)	&	(s)	&	(s)	&	(s)	&	(Intensity)	\\
\hline
56272.542903	&	2.35347	&	411.92064	&	1412.30477	&	0.00216	\\
56272.664818	&	92.02830	&	2942.31658	&	10651.19328	&	0.01879	\\
56273.280525	&	2.69546	&	529.60954	&	2353.85395	&	0.00220	\\
56274.777566	&	4.94307	&	588.46349	&	2295.00173	&	0.00274	\\
56274.878368	&	11.65929	&	529.61904	&	3648.47155	&	0.00538	\\
56281.847993	&	0.16988	&	117.69840	&	58.84445	&	0.00273	\\
56285.207831	&	2.92960	&	588.46522	&	1883.08714	&	0.00313	\\
56286.512809	&	9.49735	&	706.16448	&	5296.19386	&	0.00309	\\
56288.216910	&	1.36715	&	294.24211	&	1118.07907	&	0.00223	\\
56292.129807	&	4.13727	&	411.92410	&	2530.41322	&	0.00377	\\
56293.043839	&	6.09957	&	588.45917	&	4119.27379	&	0.00205	\\
\hline
\end{tabular}
\label{tab4}
\end{center}
\end{footnotesize}
\end{table*}

In total, 113 flares were detected from the available observations in the Kepler Mission database. The flare equivalent durations of the flare events were computed using Eq.(3) (Gershberg 1972):

\begin{center}
\begin{equation}
P = \int[(I_{flare}-I_{0})/I_{0}] dt
\end{equation}
\end{center}
where $I_{0}$ is the flux of the star in the quiet state (computed using the models of Section 2.3).

Our analysis has shown that the flare equivalent duration has a maximum for the star, independent of the flare total duration. The tests made by using the SPSS V17.0 (Green et al. 1999) and GRAHPPAD PRISM V5.02 (Dawson and Trapp 2004) programs indicated that the best function to describe dependence of the logarithm of the equivalent time duration on the total flare duration is the One Phase Exponential Association (hereafter OPEA) for the distributions of flare equivalent durations on the logarithmic scale vs. flare total durations, as it was demonstrated by Dal and Evren (2010, 2011). The OPEA function is a special one, because it has a Plateau term. The OPEA function is defined:

\begin{center}
\begin{equation}
y~=~y_{0}~+~(Plateau~-~y_{0})~\times~(1~-~e^{-k~\times~x})
\end{equation}
\end{center}
where the parameter $y$ is the flare equivalent duration on a logarithmic scale, the parameter $x$ is the flare total duration. The parameter $y_{0}$ is the shortest equivalent duration logarithm occurring in a flare for a star as described by Dal and Evren (2010). The parameter Plateau value is the upper limit for the flare equivalent duration on a logarithmic scale. Dal and Evren (2011) defined Plateau value as a saturation level for a star in the observed band. Formally, the equivalent duration approaches its saturation value for $x=\rightarrow\infty$ but for $K_{x}=ln2$ it is a halfway between its minimum and maximum values. This explains the definition of $half-life \equiv ln2/K$ (Dawson and Trapp 2004).

\begin{figure}[htb]
\includegraphics[width=1.04\textwidth]{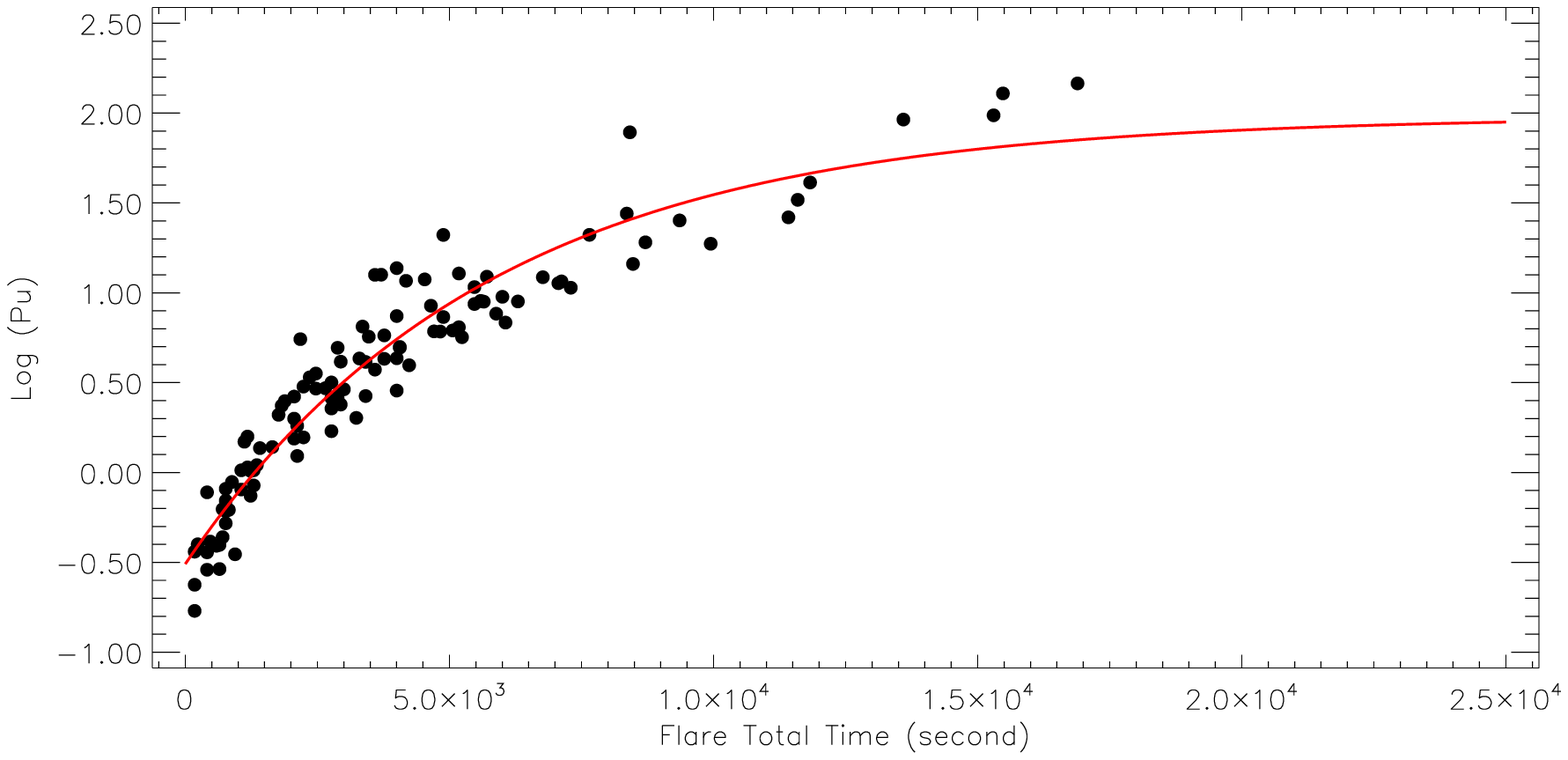}
\vspace{0.3 cm}
\FigCap{The OPEA model derived from the 113 flares detected in the observations of KIC 7885570. In the figure, the filled circles represent the observed flares, while the smooth red line.}
\label{fig:7}
\end{figure}

\setcounter{table}{4}
\begin{table*}
\begin{footnotesize}
\begin{center}
\caption{The OPEA model parameters by using the least-squares method.}
\begin{tabular}{@{}lrrc@{}}
\hline
The OPEA Best-fit Values	&		&	95$\%$ Confidence Intervals	\\
\hline					
$y_{0}$ 	&	 -0.5087$\pm$0.0485	&	 -0.6050 to -0.4125 \\
$Plateau$ 	&	 1.9815$\pm$0.1177 	&	 1.7481 to 2.2149 \\
$K$ 	&	 0.0002$\pm$0.00002	&	 0.0001 to 0.0002 \\
$Tau$ 	&	5737.9	&	 4785.0 to 7164.5 \\
$Half-life$ 	&	3977.2	&	 3316.7 to 4966.0 \\
$Span$ 	&	 2.4903$\pm$0.1048	&	 2.2823 to 2.6982 \\
\hline				
Goodness of Fit 	&	 	&	 \\
\hline				
R$^2$ 	&	 	&	 0.914 \\
$p-value$ 	&	 (D'Agostino-Pearson) 	&	 0.001 \\
$p-value $ 	&	 (Shapiro-Wilk) 	&	 0.001 \\
$p-value$ 	&	 (Kolmogorov-Smirnov) 	&	 0.001 \\
\hline
\end{tabular}
\label{tab5}
\end{center}
\end{footnotesize}
\end{table*}

The derived model is shown in Fig. 7 together with the observed flare equivalent durations. The parameters computed from the model using the least-squares method are listed in Table 5. The span value listed in Table 5 is the difference between Plateau and $y_{0}$ values.

To understand whether there are any other functions to model the distributions of flare equivalent durations on the logarithmic scale vs. flare total durations, the derived OPEA model was tested by using three different methods: the D'Agostino-Pearson normality test, the Shapiro-Wilk normality test and the Kolmogorov-Smirnov test (D'Agostino and Stephens 1986). In these tests (Table 5), the probability value called as p-value was found to be $p-value<0.001$. This means that there is no other function to model the distributions of flare equivalent durations (Motulsky 2007, Spanier and Oldham 1987).

KIC 7885570 was observed for 1301.6 d in total, from HJD 2455002.51034 to 2456304.14644 without any remarkable interruptions. The significant 113 flares were detected in the available data. The total flare equivalent duration computed from all the flares was found to be 1198.65710 s (0.33296 hours). Ishida et al. (1991) described two frequencies for the stellar flare activity. These frequencies are defined:

\begin{center}
\begin{equation}
N_{1}~=~\Sigma n_{f}~/~\Sigma T_{t}
\end{equation}
\end{center}

\begin{center}
\begin{equation}
N_{2}~=~\Sigma P~/~\Sigma T_{t}
\end{equation}
\end{center}
where $\Sigma n_{f}$ is the total flare number detected in the observations, and $\Sigma T_{t}$ is the total observing time duration, while $\Sigma P$ is the total equivalent duration obtained from all flares. In this study, $N_{1}$ frequency was found to be 0.00362 h$^{-1}$, while $N_{2}$ frequency - 0.00001.

\section{Results and Discussion}

There are several determinations of the components temperature, which vary from 5398 K (Coughlin et al. 2011) to 8254 K (Armstrong et al. 2014). We computed the temperature of the system taking the JHK magnitudes from the 2MASS All-Sky Survey Catalog (Cutri et al. 2003), and using the calibrations given by Tokunaga (2000). Firstly, the reddened colors of the system were calculated as $H-K)_{0}=0^{m}.04$ mag and $J-H)_{0}=0^{m}.23$ mag, and the temperature was found to be 6530 K. In the light curve analysis, this value, which is in agreement with the older values in the literature, was assumed as the temperature of the primary component. On the other hand, it should be noted that Armstrong et al. (2014) derived 8254 K and 8233 K for the temperatures of the primary and secondary components, respectively, using the same data taken by both the 2MASS All-Sky Survey and the Kepler Mission. The differences in the temperature values between the two studies are likely caused due to different calibrations used. However, if the temperature of the primary component was taken as 8254 K, the temperature of the secondary component would be about 7500 K. If these values were right, it would be difficult to explain how the stellar cool spot and the flare activity occurs at very high level. Thus, we assumed that the temperature of the primary component is equal 6530 K Under this assumption, the temperature of the secondary component was found to be 5732$\pm$4 K (formal fit error). The mass ratio of the component is 0.43$\pm$0.01, while the inclination: 80$^\circ$.56$\pm$0$^\circ$.01. The dimensionless potentials ($\Omega_{1}$ and $\Omega_{2}$) of the components were found to be 3.797$\pm$0.002 and 5.706$\pm$0.004, respectively, while the fractional radii of the components: 0.303$\pm$0.002 and 0.098$\pm$0.001, respectively.

Considering both the component temperatures and fractional radii found from the light curve analysis, and also the Kepler's third law, we tried to estimate the absolute parameters of the components. The masses of the components were found to be 1.59 $M_{\odot}$ and 0.98 $M_{\odot}$ for the primary and secondary, respectively. The semi-major axis ($a$) of the system is 9.68 $R_{\odot}$ and the radii of the primary and secondary components - 2.94 $R_{\odot}$ and 0.95 $R_{\odot}$, respectively. Thus, the secondary is a main sequence star, while the primary has already evolved from the main sequence. Since our investigation shows that the secondary is chromospherically active star, KIC 7885570 seems to be a RS CVn binary.

Based on derived times of eclipses an updated ephemeris was derived and the $(O-C)_{II}$ residuals were obtained. It may be seen that the $(O-C)_{II}$ residuals have two characteristic variations. First, the $(O-C)_{II}$ residuals exhibit a parabolic variation, which indicates that there is a mass loss from the system or a mass transfer from the secondary component to the primary. It is most likely an indicator of a mass loss from the system due to the flare activity occurring on the chromospherically active component. However, as it can be seen from the bottom panel of Fig. 3, if one considers the $(O-C)_{II}$ residuals of the primary and secondary minima separately, the $(O-C)_{II}$ residuals of both the primary and secondary minima vary synchronously, but in the opposite directions. This phenomenon, which indicates that the $(O-C)_{II}$ residuals of the primary and secondary minima were affected by another variation, was noticed for the first time by Tran et al. (2013) and Balaji et al. (2015). The light curve analysis revealed that the secondary component exhibits stellar spot activity. According to the results obtained by Tran et al. (2013) and Balaji et al. (2015), this explains why the $(O-C)_{II}$ residuals of both the primary and secondary minima vary synchronously in opposite directions.

Based on the out-of-eclipse variations, it is clear that one of the components is a chromospherically active star. To determine the stellar spot configurations on the active component, the data set of all the pre-whitened light curves was separated to 35 sub-data sets, and each of them was modeled by the SPOTMODEL (Rib\'{a}rik 2002; Rib\'{a}rik et al. 2003). The results indicate two spotted areas on the star at some time intervals, while there is just one spotted area some other time intervals. The parameters obtained from the SPOTMODEL analysis are listed in Table 3, while the variations of these parameters are shown in Fig. 5. There are two active spot regions located around the latitudes of $+50^\circ$ and $+90^\circ$. However, their latitudes are not stable. Nevertheless, these two regions are always separated from each other. In addition the longitudes of these areas are also rapidly varying from one cycle to another. It seems that the spotted areas migrate from one longitude to the next one with time, what indicate that there is a differential rotation on the stellar surface. In this case, it is clear why the $(O-C)_{II}$ residuals of both the primary and secondary minima vary synchronously in opposite directions. As it can be seen from the upper panel of Fig. 3 and the middle panel of Fig. 5, both the longitudinal migrations of the spotted areas and also the $(O-C)_{II}$ residuals of the primary and secondary minima are varying in the same way. However, the same behavior is seen in the variation of spot radii.

KIC 7885570 was observed for 1301.636 d. In total, 113 flares, whose parameters were computed, were detected in these data. Using these flare data, the flare frequencies $N_{1}$ and $N_{2}$ were computed as $N_{1}=$0.00362 h$^{-1}$ and $N_{2}=$0.00001. Comparing these frequencies with those computed from single UV Cet type stars, it is seen that the flare energy level found for KIC 7885570 is remarkably lower than those found for them. For example, the observed flare number per hour for UV Cet type single stars was found to be $N_{1}=$1.331 h$^{-1}$ in the case of AD Leo, and $N_{1}=$1.056 h$^{-1}$ for EV Lac. Moreover, $N_{2}$ frequency was found to be 0.088 for EQ Peg, and $N_{2}=$0.086 for AD Leo (Dal and Evren 2011). However, according to Yolda\c{s} and Dal (2016, 2017), the flare frequencies were found as $N_{1}=$0.4163 h$^{-1}$ and $N_{2}=$0.0003 for FL Lyr (B-V=0.74 mag), and $N_{1}=$0.0165 h$^{-1}$ and $N_{2}=$0.00001 for KIC 9761199 (B-V=1.303 mag). It is clearly seen that the flare frequencies of KIC 7885570 is also remarkably lower than its analogues.

Examining the parameters found from 113 flares, it was found that the flare equivalent duration on the logarithmic scale was approximately dependent on the flare total durations in a specific way. The same case was seen by Dal and Evren (2011), and they modeled this variation with the OPEA function for different stars. They also found that the Plateau values in these models change from one star to the other according to their spectral types. The Plateau value was found to be 1.9815±0.1177 for KIC 7885570. According to Dal and Evren (2011), this value is 3.014 for EV Lac (B-V=1.554 mag), 2.935 for EQ Peg (B-V=1.574 mag), and also 2.637 for V1005 Ori (B-V=1.307 mag). The maximum flare energy detected from KIC 7885570 is clearly lower than those obtained from UV Cet type single flare stars. On the other hand, Yolda\c{s} and Dal (2016, 2017) found the Plateau values of 1.232 for FL Lyr and 1.951 for KIC 9761199. In this case, the Plateau value obtained from KIC 7885570 is close to that found from KIC 9761199. The Plateau values obtained for KIC 7885570 and for KIC 9761199 have close values, but their flare frequencies are very different.

As it is seen from Table 5, the half-life value was found to be 3977.2 s, which is remarkably larger than those found from the single UV Cet type stars. For instance, it was found to be 433.10 s for DO Cep (B-V=1.604 mag), 334.30 s for EQ Peg, and 226.30 s for V1005 Ori (Dal and Evren 2011). It means that the flares can reach the maximum energy level at their Plateau value, when their total durations reach about a few$\times$5 minutes, while it requires a few$\times$66 minutes for KIC 7885570, a few$\times$39 minutes for FL Lyr and a few$\times$17 minutes for KIC 9761199 (Yolda\c{s} and Dal 2016, 2017). In addition, the maximum flare rise time ($T_{r}$) was found to be 7768.210 s for KIC 7885570. In the case of the single UV Cet type stars, the maximum flare rise time was found to be 2062 s for V1005 Ori and 1967 s for CR Dra. In the same way, the maximum flare total time ($T_{t}$) was found to be 5236 s for V1005 Ori and 4955 s for CR Dra. However, it was derived as 16890.30 s for KIC 7885570. The maximum flare rise and total times were computed as $T_{r}=$5179 s and $T_{t}=$12770.62 s for FL Lyr, and as $T_{r}=$1118.1 s and $T_{t}=$6767.72 s for KIC 9761199 (Yolda\c{s} and Dal 2016, 2017), According to these results, the flare time scales in KIC 7885570 are clearly longer than those obtained from the single UV Cet type stars. However, the flare time scales of KIC 7885570 are moderately longer than those found for FL Lyr.

In general, KIC 7885570 behaves like KIC 9761199 in terms of the Plateau parameter, and behaves like FL Lyr in terms of the flare time scales. However, in the case of the flare frequencies, KIC 7885570 is similar to none of them, because, its flare frequencies are clearly lower than those obtained for these two systems. On the other hand, this is in agreement with the results of Dal and Evren (2011). They demonstrated that the values of the parameters derived from the OPEA model depend on the B-V color of the stars. Our determination of the temperature KIC 7885570 secondary component indicates B-V=0.643 mag.

\Acknow{The authors thank O. \"{O}zdarcan for his help with the software and hardware assistance in the analyses. We also thank the referee for useful comments that have contributed to the improvement of the paper.}

\end{document}